\documentclass[prd,aps,twocolumn,a4paper,floatfix]{revtex4-1}
\usepackage{graphicx,color}

\begin{document}


\title{Critical phenomena at the threshold of immediate merger in
  binary black hole systems: the extreme mass ratio case}

\author{Carsten Gundlach$^1$, Sarp Akcay$^{1,2}$, Leor Barack$^1$ and Alessandro Nagar$^2$}
\affiliation {$^1$School of Mathematics, University of Southampton,
  Southampton SO17 1BJ, UK \\
$^2$Institut des Hautes Etudes Scientifiques, 91440 Bures-sur-Yvette, France}
\date{21 July 2012, revised 11 September 2012}


\begin{abstract}


In numerical simulations of black hole binaries,
Pretorius and Khurana [Class.~Quant.~Grav.~{\bf 24}, S83 (2007)] have
observed critical behaviour at the threshold between scattering and
immediate merger.  The number of orbits scales as $n\simeq
-\gamma\ln|p-p_*|$ along any one-parameter family of initial data such
that the threshold is at $p=p_*$. Hence they conjecture that in
ultrarelavistic collisions almost all the kinetic energy can be
converted into gravitational waves if the impact parameter is
fine-tuned to the threshold. As a toy model for the binary, they
consider the geodesic motion of a test particle in a Kerr black hole
spacetime, where the unstable circular geodesics play the role of
critical solutions, and calculate the critical exponent
$\gamma$. Here, we incorporate radiation reaction into this model
using the self-force approximation. The critical solution now evolves
adiabatically along a sequence of unstable circular geodesic orbits
under the effect of the self-force. We confirm that almost all the
initial energy and angular momentum are radiated on the critical
solution. Our calculation suggests that, even for infinite initial
energy, this happens over a finite number of orbits given by
$n_\infty\simeq 0.41/\eta$, where $\eta$ is the (small) mass ratio. We
derive expressions for the time spent on the critical solution, number
of orbits and radiated energy as functions of the initial energy and
impact parameter.


\end{abstract}


\maketitle

\tableofcontents


\section{Introduction}


Motivated by speculation about the formation of black holes in
high-energy particle collisions, there has recently been interest in
binary black hole mergers with large initial boosts; see \cite{NRHEP}
for a review. A critical surface in phase space separates initial data
which lead to merger from data which scatter at first approach, thus
defining a {\em threshold of immediate merger} in the space of initial
data. (Note that ``scatter at first approach'' here includes data
which merge at a subsequent approach.)

Pretorius and Khurana \cite{PretoriusKhurana} have numerically evolved
initial data near this threshold for equal mass, nonspinning black
hole binaries. They find that generic smooth one-parameter families of
initial data intersect the threshold once, consistent with it being a
smooth hypersurface in phase space. Moreover, they find that the
number $n$ of orbits before either merger or flying apart increases
with fine-tuning to the threshold, with $n$ scaling approximately as
\begin{equation}
\label{nscaling}
n\simeq -\gamma \ln|p-p_*|+{\rm constant},
\end{equation}
where $p$ is any {\em smooth} parameter along the family of initial
data, and $p_*$ its value at the threshold of immediate merger. The
additive constant depends on the family and on how $p$ is defined, but
the dimensionless {\em critical exponent} $\gamma$ is independent of
the family (for the reasons reviewed in Sec.~\ref{section:dynsim}). In
\cite{PretoriusKhurana}, where initial data with relative boosts of
around $0.22$ are evolved, (\ref{nscaling}) is found to be a good fit
for $1\lesssim n\lesssim 4$ with $\gamma=0.35\pm 0.03$.

In this regime, a fraction of $1$ to $1.5\%$ of the rest mass is
radiated per orbit \cite{PretoriusKhurana}. While their initial data
are rest mass-dominated, Pretorius and Khurana speculate that in
large-boost initial data fine-tuned to the critical impact parameter,
most of the initial kinetic energy, and hence most of the total
energy, can be turned into gravitational radiation over a large number
of orbits. This is quite different from the situation for either large
\cite{Martel:2003jj,Bertietal2010} or small
\cite{Davis:1971gg,OoharaNakamura1984} impact parameter. Hence in a
particle physics context, there is a small cross section
(corresponding to fine-tuned impact parameter) in which the
gravitational wave energy emitted is much larger than in a generic
collision.

In a subsequent paper using a different numerical code, Sperhake and
collaborators \cite{Sperhake} evolved initial data with boosts around
$1.5$ and $2.9$. The critical exponent $\gamma$ is estimated as $0.2$,
and the scaling law is observed for the range $0.5\lesssim n\lesssim
1.5$.  No data for critical scaling of the energy are given but in
these more relativistic collisions, the fraction of total energy
radiated is reported to be as high as $35\%$ for the higher boost.
Comparison with the results of \cite{PretoriusKhurana} suggests that
$\gamma$ and the energy radiated depend on the initial boost (initial
energy). 

Pretorius and Khurana \cite{PretoriusKhurana} note that what they
observe in full numerical relativity is similar to ``zoom-whirl''
behaviour \cite{CutlerKennefickPoisson} in the timelike geodesics of
test particles in a black-hole spacetime when the impact parameter is
fine-tuned to its critical value, now characterising the threshold of
immediate capture. Hence they propose equatorial orbits of a test
particle on a Kerr spacetime (with mass and angular momentum set to
the total mass and angular momentum of the binary) as a toy model for
the equal mass binary. They point out that the critical solutions
mediating this behaviour are the unstable circular orbits, and use
this to calculate the critical exponent for this toy model. We obtain
their results as a limiting case of ours in
Sec.~\ref{section:geodesicscaling} below.

The presence of critical phenomena suggests the existence of a
critical solution defined by having a single growing perturbation mode
and usually also characterised by a symmetry. Because energy is lost
through gravitational radiation, the critical solution cannot be
stationary, and because the black holes have rest mass it cannot be
self-similar. We suggest that the critical solution is an unstable
``as circular as possible'' orbit that in the limit where one of the
two objects becomes a test particle is indeed one of the unstable
circular orbits considered by Pretorius and Khurana.

In this paper, we extend their analysis to binaries with large but
finite mass ratio, and include radiation reaction in the self-force
approximation. We restrict to the case where the large black hole is
Schwarzschild, but our methods are in principle applicable to
Kerr. This mathematical approach allows us to clarify the nature of
the critical solution as adiabatically stationary and, based on this,
compute the orbit and radiation as a function of the initial data.

In the large mass ratio regime, we confirm the conjecture of Pretorius
and Khurana that almost all the kinetic energy of the binary (which in
turn is almost all the total energy for ultrarelativistic collisions)
is radiated if the fine-tuning of the initial data is sufficiently
good.  We find, however, that the number of orbits remains finite as
fine-tuning is improved, while (\ref{nscaling}) holds only for sufficiently
small $\eta$ and $|p-p_*|$. 

We begin by reviewing the general mathematical ideas behind critical
phenomena in dynamical systems language in
Sec.~\ref{section:dynsim}. For reference, and to establish notation,
we review timelike geodesics in Schwarzschild spacetime in
Sec.~\ref{section:Schwarzschildgeodesics}. We construct the critical
solution in the self-force approximation in
Sec.~\ref{section:selfforce}, and in Sec.~\ref{section:scaling} we
investigate its perturbations and from these we derive expressions for
the the total number of orbits and total energy radiated as functions
of the initial energy and impact parameter.
Sec.~\ref{section:conclusions} summarises our results and gives an
outlook on the comparable mass case.

Throughout the paper $\simeq$ denotes equality up to subleading terms,
and $\sim$ equality up to subleading terms and an overall constant
factor, and $:=$ denotes a definition. 


\section{Dynamical systems ideas}
\label{section:dynsim}


We briefly review {\em type~I critical phenomena} and the calculation
of the related critical exponent, in the language of abstract
dynamical systems. Let $\phi$ represent a point in phase space. This
could be either a vector of variables or a vector of fields at one
moment of time. Let $\phi(t)$ be a trajectory in phase space that
represents a solution. For the basic critical phenomena picture, we do
not need to distinguish between field theories and finite-dimensional
dynamical systems (and so do not write any $x$-dependence). This will
later allow us to approximate a field theory (general relativity) by a
finite-dimensional dissipative dynamical system (particle orbits with
self-force).

Assume that at late times there are two qualitatively distinct
outcomes, such as scattering and plunge in our system. There must then
be a hypersurface in phase space separating these two basins of
attraction, called the {\em critical surface}. As solution curves in
phase space cannot intersect, the critical surface must be a dynamical
system in its own right. Assume that the critical surface, considered
as a dynamical system, has an attractor, called the {\em critical
  solution} $\phi_*$. As $\phi_*$ is a fixed point (independent of
$t$), its linear perturbations must be a sum of modes of the form
$e^{\lambda_i t}\phi_i$, with $\phi_i$ also independent of $t$. As
$\phi_*$ is an attractor within the critical hypersurface, which
itself is a repeller, $\phi_*$ must have precisely one growing mode,
pointing out of the critical surface, for some (real) $\lambda_0>0$ \cite{LRR}.

Consider a one-parameter family of initial data, with parameter $p$,
such that this family intersects the critical surface at $p=p_*$. The
evolution of the precisely critical initial data with $p_*$ must lie
in the critical surface and hence will find $\phi_*$. Consider now
initial data near the critical surface, with $p-p_*$ sufficiently
small so that $\phi(t)$ passes close to $\phi_*$ during the evolution,
and we can use perturbation theory about $\phi_*$. There is then a
range of $t$ where the decaying modes can be already neglected and the
growing mode $\phi_0$ is still small enough for perturbation theory to
hold, leading to the approximation
\begin{equation}
\label{C1def}
\phi(t)\simeq \phi_* + C_1(p-p_*)e^{\lambda_0 t}\phi_0,
\end{equation}
where $C_1$ is some constant that depends on the particular
one-parameter family. Now define $t_*$ by
\begin{equation}
\label{C2def}
\phi(t_*)\simeq \phi_* + C_2\phi_0,
\end{equation}
where $C_2$ is a fairly arbitrary constant, representing a reference deviation
from the critical solution where it becomes apparent on which side of
the critical surface the solution is going to end up. One then finds
\begin{equation}
\label{C3def}
t_*\simeq -\Gamma \ln|p-p_*|+C_3,
\end{equation}
where $C_3$ depends on $C_1$ and $C_2$, and $\Gamma=1/\lambda_0$ is
called the {\em critical exponent}. 

Essentially the same picture holds if $\phi_*$ is not a fixed point
but a limit cycle (periodic in $t$). Then $\phi_0$ is also periodic in
$t$ and so is $C_3$, and this leads to a modulation of the scaling law
periodic in $\ln|p-p_*|$. This generalisation is not relevant for our
application, but might be for extreme-mass ratio orbits in Kerr not
in the equatorial plane, or more generally for binaries with spin.

However, another generalisation is relevant here. As we shall see, in
the geodesic toy model of Pretorius and Khurana the critical solution
is not an isolated attractor, but instead there is a line of critical
solutions (each of which represents an unstable circular orbit, and
has precisely one growing mode). In the self-force approximation to
extreme mass ratio binaries, by contrast, these merge into a
single critical solution, which can be approximated as an adiabatic
motion along the line of unstable circular geodesic orbits, and so is
slowly time-dependent. The general dynamical system picture above
remains again basically unchanged, except that in the first case
$\phi_*$, $\phi_0$ and $\lambda_0$ form a one-parameter family
(parameterised for example by the energy of the orbit), and in the
second case they become slowly varying functions of $t$. 


\section{Test particles on Schwarzschild spacetime}
\label{section:Schwarzschildgeodesics}


\subsection{Equations of motion}
\label{section:geodesicEOM}


To establish notation, we review the trajectories of test particles,
modelled as geodesics, on Schwarzschild spacetime. Let
$t,r,\theta,\varphi$ be the usual Schwarzschild coordinates, so that
the Schwarzschild metric is
\begin{eqnarray}
ds^2&=&-\left(1-{2\over r}\right)dt^2+\left(1-{2\over
  r}\right)^{-1}dr^2 \nonumber \\ && +r^2(d\theta^2+\sin^2\theta d\varphi^2),
\end{eqnarray}
and let a dot denote the derivative with respect to the proper time
$\tau$ of the particle. Throughout this paper we use units such that
Newton's constant $G$, the speed of light $c$ and the mass $M$ of the
Schwarzschild spacetime are all unity.  Let $u^a=\dot x^a=dx^a/d\tau$
be the 4-velocity of the particle. Without loss of generality let the
orbit be in the plane $\theta=\pi/2$.  Define
\begin{eqnarray}
\label{Edef}
E&:=&-\left({\partial\over\partial t}\right)^a u_a, \\
\label{Ldef}
L&:=&\left({\partial\over\partial \varphi}\right)^a u_a,
\end{eqnarray}
representing the energy and angular momentum {\em per rest mass} of
the test particle. Retaining the convention $c=G=1$, but restoring
$M$, and with $m$ the mass of the test particle, its physical energy
and angular momentum are $m E$ and $m ML$. As $\partial/\partial t$
and $\partial/\partial \varphi$ are Killing vectors of the
Schwarzschild spacetime, $E$ and $L$ are conserved quantities in the
sense that
\begin{eqnarray}
\label{Econstant}
\dot E&=&0, \\
\label{Lconstant}
\dot L&=&0
\end{eqnarray}
along geodesic orbits. Hence we obtain
\begin{eqnarray}
\label{Edefbis}
\dot t&=&\left(1-{2\over r}\right)^{-1}E, \\
\label{Ldefbis}
\dot\varphi&=& {L\over r^2}.
\end{eqnarray}
By definition, $E>0$ for any timelike geodesic for $r>2$, where
$\partial/\partial t$ is future timelike. Without loss of generality
we shall also assume $L\ge 0$. The normalisation condition $u_au^a=-1$ can
then be expressed as
\begin{equation}
\label{Edef3}
E^2=\dot r^2+V(L,r), 
\end{equation}
where 
\begin{equation}
\label{Vdef}
V(L,r):=\left(1-{2\over r}\right)\left(1+{L^2\over r^2}\right).
\end{equation}
Geodesics on Schwarzschild obey
\begin{equation}
\label{geodesics}
u^b\nabla_b u^a=\ddot x^a+\Gamma^a{}_{bc}\dot x^b \dot x^c=0,
\end{equation}
but (\ref{Econstant},\ref{Lconstant}) together with the normalisation
condition (\ref{Edef3}) can be used to reduce this 6-dimensional
dynamical system to a 3-dimensional one in the variables $(r,\dot r,
L)$, with the equations of motion
\begin{eqnarray}
\label{ddotreqn}
\ddot r &=& -{1\over 2}V_{,r}(L,r),\\
\label{dotLeqn}
\dot L &=& 0,
\end{eqnarray}
and where $E(r,\dot r,L)$ defined by (\ref{Edef3}) is an integral of
the motion. (Here and in the following, commas denote partial
derivatives.) Alternatively, as long as $\dot r$ does not change sign,
the system can be written in the variables $(r,L,E)$ in the form
\begin{eqnarray}
\label{rdot}
\dot r &=& \pm\sqrt{E^2-V(L,r)}, \\
\dot L &=& 0, \\
\label{Edot}
\dot E &=& 0,
\end{eqnarray}
which now shows both integrals of the motion explicitly. In either case,
the variables $(t,\varphi)$ are evolved by
(\ref{Edefbis},\ref{Ldefbis}) and play no dynamical role, while the
remaining component of (\ref{geodesics}) is consistent with our assumption
$\theta=0$.

\begin{figure}
\includegraphics[width=8cm]{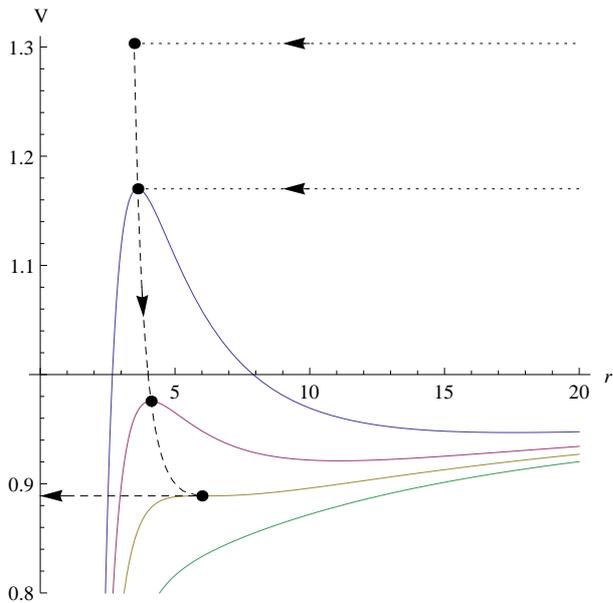}
\caption{ {\bf For Sec.~\ref{section:geodesicEOM}:} Illustration of
  the effective potential (\ref{Vdef}) for different values of
  $L$. From above, the four full lines illustrate the cases $ L>4$,
  $\sqrt{12}< L<4$, $ L=\sqrt{12}$ and $ L<\sqrt{12}$.  Note that all
  these curves approach $V\to 1$ as $r\to \infty$. --- {\bf For
    Sec.~\ref{section:qualitative}:} The dashed line represents the
  critical solution. Its almost vertical segment represents the
  adiabatic phase. As illustrated by the beads, it is given by
  $r(t)=r_-(L(t)),E^2(t)=V_-(L(t))$, with $\dot L(t)$ determined by
  radiation. The horizontal segment represents the plunge phase.  The
  two dotted lines represent orbits coming in from infinity at the
  critical impact parameter for different $E$, and asymptoting to the
  critical solution. Radiation reaction has been neglected in the
  approach and plunge phases, while $\dot r$ has been approximated as
  zero in the adiabatic phase. With radiation reaction, to order
  $\eta$ the approach and plunge phase curves bend slightly downwards,
  the adiabatic curve is slightly further to the right, and the kinks
  are slightly smoothed out. [The beads are the maxima of the
    potential for $ L=5,4.6,3.9,\sqrt{12}$, and the lowest curve
    (without a maximum) corresponds to $ L=3$.]
\label{figure:geodesicpotential}}
\end{figure}

\begin{figure}
\includegraphics[width=8cm]{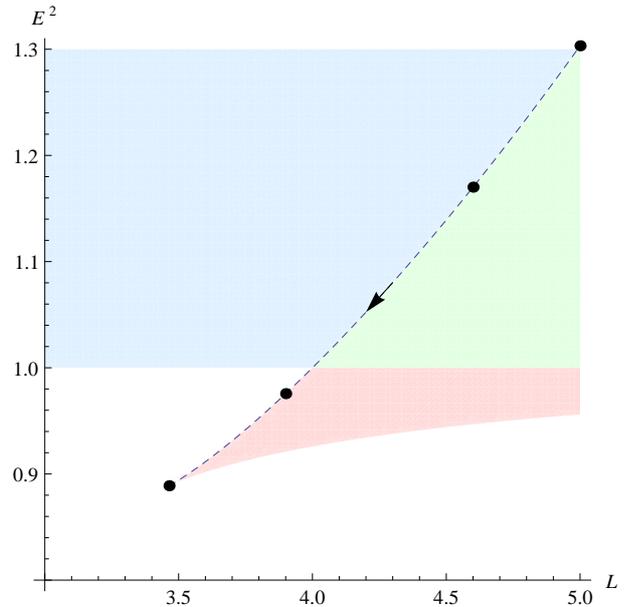}
\caption{ {\bf For Sec.~\ref{section:Schwarzschildcritical}:} The
  space of geodesics on Schwarzschild, labelled by angular
  momentum per unit mass $L$ and energy per unit mass (squared) $E^2$
  of the particle. The bottom right shaded region contains bound
  orbits that oscillate between a minimum and maximum radius. Its
  lower edge is given by stable circular orbits. The left shaded
  region contains orbits that come in from infinity with low impact
  parameter and plunge. The top right shaded region contains orbits
  that come in from infinity with high impact parameter and
  scatter. The line dividing these two regions is therefore the
  threshold of immediate capture, given by $E^2=V_-(L)$ (which
  characterises unstable circular orbits). The white region
  corresponds to orbits that emerge from the white hole horizon and
  plunge back into the black hole, and which are not of interest to us
  here. --- {\bf For Sec.~\ref{section:qualitative}:} In the
  self-force approximation, the critical solution moves adiabatically
  along the line of unstable circular geodesic orbits, as indicated by
  the dashed line and arrow. The beads and arrow are as in
  Fig.~\ref{figure:geodesicpotential}. (The approach and plunge phases
  seen in Fig.~\ref{figure:geodesicpotential} are essentially in the
  $r$ direction, which is suppressed here.)
\label{figure:geodesicphasespace}}
\end{figure}

The shape of the effective potential $V(L,r)$ for different ranges of
$L$ is illustrated in Fig.~\ref{figure:geodesicpotential}. (The
dashed and dotted curves should be ignored for now.) As
$V(L,\infty)=1$ for any $L$, orbits with $E\ge 1$ are unbound.  For
$L>\sqrt{12}$, the effective potential has a maximum at $r=r_-$
and a minimum at $r=r_+$, given by
\begin{equation}
\label{rpm}
r_\pm:={ L( L\pm L_{12})\over 2},
\end{equation}
where it takes values 
\begin{equation}
\label{Vpm}
V_\pm:=V(r_\pm)={2(2 L\pm L_{12})^2\over 9 L( L\pm
   L_{12})}.
\end{equation}
Here we have introduced the shorthand
\begin{equation}
L_{12} := \sqrt{ L^2-12}.
\end{equation}
The minimum and maximum give rise to a stable and an unstable circular
orbit. These merge for $L=\sqrt{12}$, at $r_+=r_-=6$.  

For trajectories coming in from infinity, and hence with $E\ge 1$, a
natural parameter of the initial data is the impact parameter $b>0$,
defined by
\begin{equation}
b:=\lim_{r\to\infty}r\sin|\varphi(r)-\varphi(\infty)|.
\end{equation}
With
\begin{equation}
{d\varphi\over dr}={L\over r^2\sqrt{E^2-V(L,r)}},
\end{equation}
we find 
\begin{equation}
\label{bLE}
E^2-1={ L^2\over b^2}.
\end{equation}

We define the (instantaneous) orbital angular frequency
\begin{equation}
\label{Omegadef}
\Omega:={d\varphi\over dt}={\dot\varphi\over\dot t}
=\left(1-{2\over r}\right){L\over r^2E}
={V_{,L}\over 2E}.
\end{equation}
For both the stable and unstable circular
geodesic orbits, we have
\begin{eqnarray}
\label{Ec}
E&=&{r-2\over\sqrt{r(r-3)}}, \\
\label{Lc}
L&=&{ r\over\sqrt{r-3}}, \\
\label{Omegac}
\Omega &=& r^{-3/2}
\end{eqnarray}
We will be particularly
interested in the ultrarelativistic limit $E\gg 1$, where
\begin{eqnarray}
\label{Eultra}
E&\simeq&{1\over\sqrt{3}}( r-3)^{-1/2}, \\
\label{Lultra}
 L&\simeq&3( r-3)^{-1/2}\simeq 3\sqrt{3} E, \\
\label{Omegaultra}
\Omega&\simeq& {1\over 3\sqrt{3}}.
\end{eqnarray}


\subsection{Critical solution}
\label{section:Schwarzschildcritical}


For simplicity of presentation, in the following we discuss only
geodesic orbits with $E>1$ and which initially have $\dot r<0$ and
$r>r_-(L)$, coming in from large $r$. If the impact parameter is
fine-tuned so that $E^2=V_-(L)$, the orbit asymptotically approaches the
unstable circular orbit $r=r_-(L)$, without ever plunging or moving
back out. Orbits with the same $L$ but smaller $E$ scatter, while
orbits with the same $L$ but larger $E$ plunge. Hence the critical
surface in the space of initial data parameterised by $(r,E,L)$ with
$\dot r<0$ is formed by the surface $E^2=V_-(L)$, $r>r_-(L)$. The
space of geodesic orbits is illustrated in
Fig.~\ref{figure:geodesicphasespace}.

We define the critical impact parameter $ b_*(E)$ by
$E^2=V_-(L=L(E, b_*))$ and find
\begin{equation}
\label{b*E}
b_*(E)={\sqrt{27E^4-36E^2+8+E(9E^2-8)^{3/2}}\over \sqrt{2} (E^2-1)},
\end{equation}
which has the well-known limits
\begin{equation}
\label{bstarultra}
b_*(\infty)=3\sqrt{3}
\end{equation}
and $b_*(1)=\infty$ (every particle released from rest at infinity
falls into the black hole).


\subsection{Perturbations of the critical solution}


The linearisation of the dynamical system
(\ref{ddotreqn},\ref{dotLeqn}) about the critical solution
  $r=r_-(L)$ is
\begin{eqnarray}
\label{ddotdeltar}
\delta \ddot r&=&-{1\over 2}V_{,rr-}\delta r-{1\over
  2}V_{,rL-}\delta L,\\ 
\label{dotdeltaL}
\delta \dot L &=&0,
\end{eqnarray}
where $V_{,rr-}$ and $V_{,rL-}$ are $V_{,rr}$ and $V_{,rL}$ evaluated
on the critical solution $r=r_-(L)$. 
We can decouple these equations as
\begin{eqnarray}
\label{ddotxbis}
\ddot x&=&\Gamma^{-2}x, \\
\dot\delta L &=&0,
\end{eqnarray}
where
\begin{eqnarray}
\label{xdef} 
x&:=&\delta r+{V_{,rL-}\over V_{,rr-}}\delta L, \\
\label{GammaV}
\Gamma&:=&\left(-{1\over 2}V_{,rr-}\right)^{-1/2}
={ L^{3/2}( L- L_{12})^2\over 4  L_{12}^{1/2}}.
\end{eqnarray}
With $E^2=V_-(L)$ on the critical solution, we can consider $\Gamma$
also as function of $E$, but we have not been able to express
$\Gamma(E)$ in closed form.

From (\ref{ddotdeltar},\ref{dotdeltaL}) we see that the critical solution
has exponentially growing and decaying modes $x=\exp(\pm\tau/\Gamma)$
with $\delta L=0$, and a neutral mode $\delta L=1$ with $x=0$. Hence
it has precisely one growing mode, as required of a critical
solution. However, in the geodesic approximation we do not have a
critical solution which is a global attractor within the critical
surface, but rather a line of critical solutions, to which the neutral
mode is tangent.


\section{Critical solution
in the self-force approximation}
\label{section:selfforce}


\subsection{Qualitative discussion}
\label{section:qualitative}


Our aim is to replace the model of the binary merger as a test
particle on Schwarzschild spacetime with a more realistic model that
in particular incorporates radiation reaction, but which still
approximates the Einstein equations as a finite-dimensional dynamical
system by integrating out the radiation. 

One such model is the extreme mass-ratio approximation, where the loss
of energy and angular momentum are calculated to leading order in
$\eta:=m/M$, where $M=1$ is the mass of a large black hole and $m$
that of the smaller object, modelled as a point particle. We shall use
an adiabatic approximation, where the particle is considered to move
on a geodesic with $E$ and $L$ now slowly decaying, and a force term
appearing in consequence on the right-hand side of the geodesic
equation. We can then still use the effective potential picture to
write the equations of motion, where $E(t)$ and $L(t)$ now evolve
under the effect of the radiation reaction.

In short, the critical solution sits approximately at the maximum of
the potential, with $E(t)^2=V_-(L(t))+O(\eta)$ and $r(t)=
r_-(L(t))+O(\eta)$, while that maximum itself evolves under radiation
reaction, starting with the ultrarelativistic limit (\ref{Lultra}) as
$r\simeq 3$ and $E\to\infty$, and ending with a plunge from $E\simeq
8/9$, $L\simeq\sqrt{12}$ and $r\simeq 6$ (see
Fig.~\ref{figure:geodesicpotential}). Our self-force calculation
suggests that infinite $E$ and $L$ occur at {\em finite} $\tau$, $t$
and $\varphi$ in the past, but as we discuss below, we cannot
consistently reach $E=\infty$ within the self-force approximation.

Any initial data that are perfectly fine-tuned to the threshold of
immediate merger, with arbitrarily large $E$, join the critical
solution, which acts as an attractor in the critical surface. This is
indicated schematically in Figs.~\ref{figure:geodesicpotential} and
\ref{figure:geodesicphasespace}, where the dotted lines with arrows
indicate two fine-tuned solutions approaching the critical solution
(dashed) and remaining on it until and through the plunge. Note that
the line of critical fixed points in the test particle picture has
become a single, adiabatically evolving critical solution.


\subsection{Equations of motion}


In the extreme mass ratio limit, the larger object in the binary is
approximated as a Schwarzschild black hole, and the lighter one as a
particle moving in this background spacetime, with the equation of motion
\begin{equation}
\label{udot}
u^b\nabla_b u^a=F^a,
\end{equation}
where $u^b\nabla_b u^a=0$ describes a geodesic on the background
Schwarzschild spacetime, and $F^a$ is the self-force per particle rest
mass, which is proportional to the mass ratio $\eta$ to leading
order. (We use terminology customary in the self-force literature
where it is really $mE$, $mML$ and $mF^a$ that have dimensions of
energy, angular momentum and force.) $\nabla_a$ is the covariant
derivative in the background spacetime, and indices are moved
implicitly with the background metric.

As before, we define $E$ and $L$ by (\ref{Edefbis}) and
(\ref{Ldefbis}), and we impose the normalisation $u^au_a=-1$ in the
form (\ref{Edef3}). The time derivative of $u^au_a=-1$ gives the usual
constraint on any 4-force, namely
\begin{equation}
\label{uF}
u_a F^a=0.
\end{equation}
From (\ref{udot}), we now have the 3-dimensional dynamical system
\begin{eqnarray}
\label{vdotbis}
\ddot r &=& -{1\over 2}V_{,r}(L,r) +F^r,\\
\label{Ldotbis}
\dot L &=& F_\varphi.
\end{eqnarray}
$E$ can be read off as $E(r,\dot r,L)$ from 
(\ref{Edef3}), or can be evolved using the third component of
(\ref{udot}), which can be written as
\begin{equation}
\dot E = - F_t.
\end{equation}
These are compatible because of (\ref{uF}), which can be written as
\begin{equation}
\label{uFbis}
\dot rF^r+E(F_t+\Omega F_\varphi)=0,
\end{equation}
where $\Omega$ is defined by (\ref{Omegadef}). $\varphi$ and $t$
  are evolved via (\ref{Ldefbis}) and (\ref{Edefbis}).


\subsection{Adiabatic expansion}


By definition, the critical solution is the one that is balanced
between plunging and running out to infinity. In the absence of
radiation reaction, this clearly means it is a circular geodesic
sitting on top of the potential. In the presence of radiation
reaction, this definition becomes teleological: the critical solution
hesitates as long as possible between plunging and running out to
infinity. In practice we use adiabaticity to define the critical
solution as being as circular or as stationary as possible: $r$ should
vary over a radiation reaction timescale, that is, $\dot r$ should be
proportional to the mass ratio $\eta$ to leading order in the
self-force expansion.

We formalize the adiabatic expansion through the series ansatz
(slow-time expansion) \cite{HindererFlanagan}
\begin{eqnarray}
\label{Fk}
 F^a_*(\tau)&=&\sum_{k=1}^\infty \eta^k F^a_{k}(\hat\tau), \\
\label{rk}
r_*(\tau)&=&\sum_{k=0}^\infty \eta^k r_{k}(\hat\tau), \\
\label{Lk}
L_*(\tau)&=&\sum_{k=0}^\infty \eta^k L_{k}(\hat\tau), \\
\label{Ek}
E_*(\tau)&=&\sum_{k=0}^\infty \eta^k E_{k}(\hat\tau),
\end{eqnarray}
where we have defined the slow time
\begin{equation}
\hat\tau:=\eta\tau, 
\end{equation}
the suffix $*$ denotes the critical solution, and the normalisation
condition (\ref{Edef3}) is obeyed order by order by virtue of
(\ref{uFbis}).  To leading order,
\begin{eqnarray}
 F^a_*&=&\eta F^a_1+O(\eta^2), \\
\ddot r_*&=&\eta^2 r_0''+O(\eta^3), \\
\dot L_*&=&\eta L_0'+O(\eta^2), \\
\dot E_*&=&\eta E_0'+O(\eta^2). 
\end{eqnarray}
where $\dot f:=df/d\tau$ as before, and $f':=df/d\hat\tau$. 
Substituting this into (\ref{vdotbis},\ref{Ldotbis}), we obtain to
leading order [$O(1)$ in (\ref{vdotbis}), and $O(\eta)$ in (\ref{Ldotbis})]
\begin{eqnarray}
0 &=& V_{,r}(L_0,r_0)  \quad \Rightarrow \quad r_0 = r_-(L_0), \\
\label{L0'}
L_0'&=&F_{\varphi 1}, 
\end{eqnarray}
and this allows us to integrate $r_0$ and $L_0$.  Hence,
$F^a_1(\hat\tau)=F^a_1(r_0(\hat\tau))$ is the first-order self-force
on a circular geodesic orbit of radius $r_0$, which is known
\cite{Barack:2007tm}.  To next order,
\begin{eqnarray}
\label{Fr1eqn}
0&=&-{1\over 2}[V_{,rr-}(L_0,r_0)r_1+V_{,rL}(L_0,r_0)L_1]+F^r_1, \\
L_1'&=&F_{\varphi 2}
\end{eqnarray}
formally give us $r_1$ and $L_1$, but we cannot use this to calculate
the critical solution beyond $(r_0,L_0)$. $F^a_2$ is the second-order
self-force. It comprises terms that are quadratic in metric
perturbations and corrections due to the fact that the critical
solution is not exactly geodesic. Both have not yet been
calculated. Note, however, that (\ref{Fr1eqn}) can be rewritten as
\begin{equation}
0=-{1\over 2}V_{,r}(L_0+\eta L_1,r_0+\eta r_1)+\eta F^r_1+O(\eta^2),
\end{equation}
and so its physical meaning is that the critical solution is slightly
off the peak of the geodesic potential, held there (to this order) by
the conservative part $F^r_1$ of the self-force. 


\subsection{Self-force input}


To leading order, our analysis requires knowledge of
$\dot{L}=F_\varphi$ along circular orbits with $3<r\le 6$. In practice,
we calculate $\dot E=-F_t$, which on circular orbits is related to
$\dot L$ by $\dot L=\dot E/\Omega$.  To obtain $\dot E$ we used the
self-force code developed in Ref.\ \cite{Akcay:2010dx}, which is a
frequency-domain variant of the original Barack-Sago code
\cite{Barack:2007tm}. The code was originally developed to deal with
stable orbits, but it can handle unstable (circular) orbits
without any further development. The code takes as input the orbital
radius $r$ and returns the value of $\dot E$ along that orbit. We have
obtained accurate data for a dense sample of $r$ values in the range
$3<r\le 6$. The numerical data and details of their derivation are given in
Appendix~\ref{appendix:data}.

Our numerical results can be approximated by the semi-heuristic
fitting formula
\begin{equation}
\label{Edotfit}
\dot E\simeq-\eta r^{-5} \ z^{-\alpha}\ (a_0+a_1z+a_2z^2+a_3z^3),
\end{equation}
where $z:=1-3/ r$ and the parameters
\begin{eqnarray}
&&\alpha\simeq 1.773163, \quad a_0\simeq 2.87683, \quad a_1\simeq
-4.01414, \nonumber \\  &&a_2\simeq 10.5371, \quad a_3\simeq -3.79087
\label{fitvals}
\end{eqnarray}
have been determined by a least-squares fit of $\dot E$ versus
$z$. The $r^{-5}$ decay at large $r$ is well-known. We have modelled
the blow-up of the self-force at the light ring by a single 
power $z^{-\alpha}$, although we have no rigorous theoretical argument
for this. The numerical data we have fitted to are also not very close
to the light ring, with the smallest value of $r$ at $3.02$,
corresponding to $E\simeq 4.15$.

\begin{figure}
\includegraphics[width=8cm]{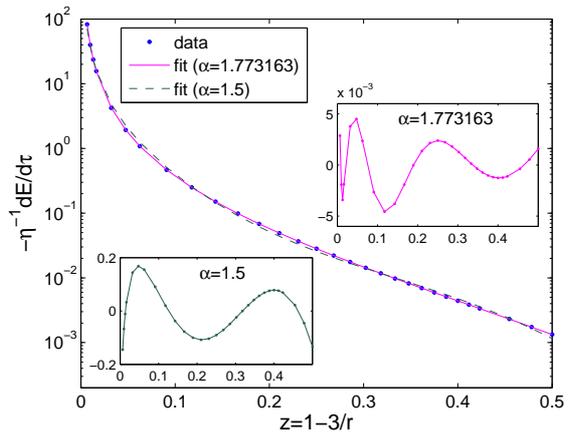}
\caption{Energy loss $\dot E/\eta$ as a function of $z:=1-3/r$, for a
  circular geodesic orbit in the self-force approximation, and two
  closed-form fits. The fitting ansatz is given by Eq.~(\ref{Edotfit})
  with $\alpha$ either fitted (solid line), or fixed at $\alpha=3/2$
  (dashed line). The self-force data (points) are given in
  Appendix~\ref{appendix:data}. The insets show the relative
  differences between the fit and the data for the two models (note
  the different vertical scales).}
\label{figure:dEdtau}
\end{figure}

A tentative theoretical argument for determining $\alpha$ is that at
sufficiently high energy the metric perturbation should be
proportional to the total energy $E$ of the particle and therefore the
fluxes as seen by an observer at rest should be proportional to $E^2$,
so that
\begin{equation}
\dot E\sim \dot t\, E^2 \sim z^{-3/2}.
\end{equation}
[Note that, from (\ref{Edefbis}) and (\ref{Ec}), $\dot t=z^{-1/2}$ on
  circular geodesics.] A least-squares fit to (\ref{Edotfit}) where
$\alpha=1.5$ is held fixed results in a maximum fitting error of
$17\%$. By comparison, the maximum fitting error with (\ref{fitvals})
is $0.5\%$, which is much smaller and comparable with the maximal
estimated error of the self-force calculation (see
Appendix~\ref{appendix:data}). The numerical self-force results and
both fits are shown in Fig.~\ref{figure:dEdtau}. For our computations
and plots in the remainder of the paper, we shall use the fit
(\ref{Edotfit}) with (\ref{fitvals}) for the range $r\ge 3.02$ for
which we have self-force data, corresponding to $E\lesssim 4.15$. 

\begin{figure}
\includegraphics[width=8cm]{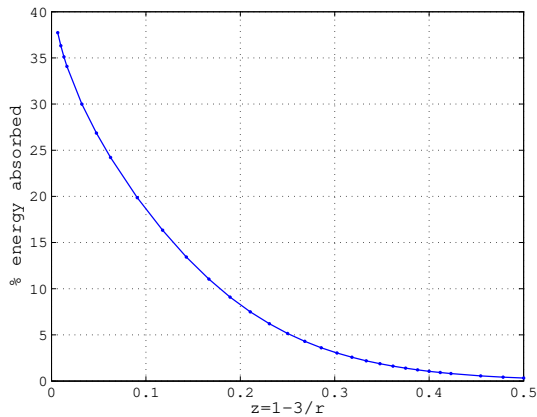}
\caption{The {\em fraction} of the total energy loss $\dot E/\eta$
  that is absorbed by the large black hole, as a function of
  $z:=1-3/r$. This is based on the data in
  Appendix~\ref{appendix:data}. Extrapolation gives a fraction of
  $\simeq 42\%$ at the light ring itself.}
\label{figure:absorption}
\end{figure}

Fig.~\ref{figure:absorption} shows the energy absorbed by the large
black hole as a fraction of the total energy loss (radiated to
infinity plus absorbed by the large black hole). This figure is
relevant for computing the total energy radiated to infinity, but not
for our construction of the critical solution, where only the total
energy loss matters. 

\subsection{Integration of the critical solution}


In the presence of radiation reaction, for a given mass ratio $\eta$,
there is a single critical solution where $E$, $L$, $\Omega$, $r$,
$\tau$ and $\varphi$ are all monotonous functions of one another, with
$E$ and $L$ decreasing from $\infty$ to finite values at plunge, $r$
increasing from $3$ to $6$, and $\Omega$ decreasing from $3^{-3/2}$ to
$6^{-3/2}$, all as $\tau$ and $\varphi$ increase. In this paper we
construct the critical solution only to leading order in $\eta$. Then
$E$, $L$ and $\Omega$ and $r$ are known functions of one
another, related by (\ref{Ec}-\ref{Omegac}). Furthermore, the
$\eta$-dependence of the critical solution is of the simple form
$r=r(\hat\tau)=r(\hat\varphi)$, where we define the ``slow angle''
\begin{equation}
\hat\varphi:=\eta\varphi
\end{equation}
in analogy to the slow time $\hat\tau$. Finally, the ODEs obeyed by
$\hat\tau(r)$ and $\hat\varphi(r)$ are separable, and can be solved
by integration.

\begin{figure}
\includegraphics[width=8cm]{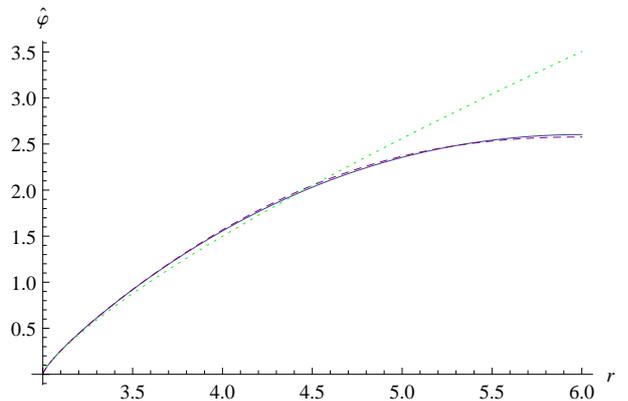}
\caption{Slow angle $\hat\varphi$ as a function of radius $r$ in the
  critical solution. We have fixed $\hat\varphi=0$ at $r=3$. The full
  line represents a numerical integration of (\ref{dphidr}), with
  $\dot E(r)$ given by the fitting formula (\ref{Edotfit}).  The
  dashed line represents the closed form approximation discussed in
  the text below Eq.~(\ref{dphidr}), and the dotted line the
  ultrarelativistic approximation (\ref{phiofrapprox}). Note that here
  and in the following figure, the narrow section $r<3<3.02$ of the domain
  is derived from an extrapolation of our self-force data.}
\label{figure:phi_vs_r}
\end{figure}

We obtain ${\hat\varphi}(r)$ by (numerical) integration of 
\begin{eqnarray}
{d{\hat\varphi}\over d r}&=&\eta {d\varphi\over dt}{dt\over
  d\tau}{d\tau\over dE}{dE\over dr} 
= \Omega \dot t {dE\over dr}{\eta\over\dot E} \nonumber \\ &=&
{1\over 2} r^{-5/2}(r-3)^{-2}(r-6){\eta\over\dot
  E},
\label{dphidr}
\end{eqnarray}
after substituting (\ref{Edotfit}). A very good closed form
approximation to this integral in terms of elementary functions can be
obtained by approximating $(d\hat\varphi/dr)/(r-6)$ as a series in
$(r-3)$ with three terms, multiplying back by $(r-6)$, and integrating
this. The full result, the closed form approximation, and the
ultrarelativistic approximation discussed below are shown in
Fig.~\ref{figure:phi_vs_r}. For fixed $\eta$, this graph gives us the
{\em shape} of the critical orbit.

We only have reliable self-force data for $r\ge 3.02$, corresponding
to $E\lesssim 4.15$, but we are interested in the critical solution up
to $E=\infty$. We have therefore extrapolated our fit (\ref{Edotfit})
with (\ref{fitvals}) down to $r=3$ in producing
Fig.~\ref{figure:phi_vs_r}, and have used this extrapolation to set
$\hat\varphi(3)=0$ at $E=\infty$ as a matter of convention. Note,
however, that almost the entire range of $\hat\varphi$ in this plot
occurs for $r>3.02$, where we do have reliable self-force data, with
$\hat\varphi(3.02)\simeq 0.07$ compared to $\hat\varphi(6)\simeq
2.6$. This qualitative feature is robust, requiring only that $\dot E$
diverges at the light ring faster than $(r-3)^{-1}$, that is
$\alpha>1$. Our self-force data for $r\le 3.02$ support this
qualitative feature robustly, with $\alpha\simeq 1.77$, while our
crude theoretical model suggests $\alpha=3/2$.

Hence, the total number of orbits in the critical solution, from
infinite energy down to plunge, is given by
\begin{equation}
\label{totalorbits}
n_{\rm tot}={\hat\varphi(6)\over 2\pi\eta}\simeq{0.41\over \eta},
\end{equation}
and we expect the numerical prefactor to depend only weakly on our
extrapolation of the self-force data.

\begin{figure}
\includegraphics[width=8cm]{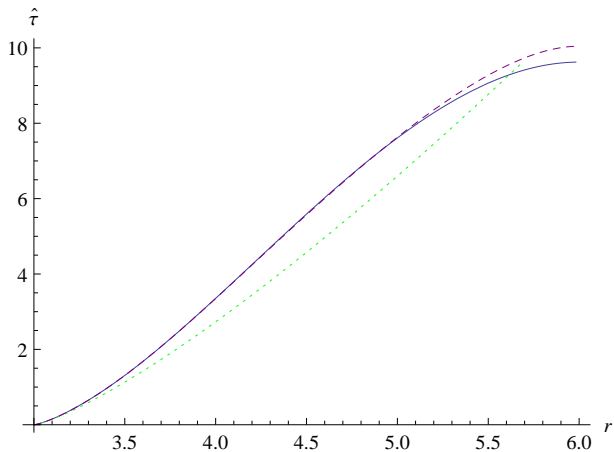}
\caption{Slow time $\hat\tau$ as a function of radius $r$ in
    the critical solution. We have fixed $\hat\tau=0$ at $r=3$. The
    full line represents a numerical integration of
    (\ref{drdtau}). The dashed line represents the closed form
    approximation for $\hat\tau(r)$ discussed in the text below
    Eq.~(\ref{drdtau}) and the dotted line the ultrarelativistic
    approximation (\ref{tauofrapprox}).}
\label{figure:r_vs_tau}
\end{figure}

We obtain $\hat\tau(r)$ by (numerical) integration of
\begin{equation}
\label{drdtau}
{d\hat\tau\over dr} ={\eta\over\dot E} {dE\over dr}.
\end{equation}
A reasonably good closed form approximation for $\hat\tau(r)$ can be
obtained by expanding $(d\hat\tau/dr)/(r-6)$ into a series in $(r-3)$
with four terms, multiplying back by $(r-6)$, and integrating. The
full result and its approximations are shown in
Fig.~\ref{figure:r_vs_tau}. Our self-force data suggest that
$\hat\tau$, too, is finite at infinite energy, and we use the
convention that $\hat\tau=0$ at $E=\infty$. [From
  (\ref{drdtau},\ref{Eultra}), the precise criterion is that
  $(r-3)^{-3/2}/\dot E$ is integrable.]  Again, compare
$\hat\tau(3.02)\simeq 0.02$ with $\hat\tau(6)\simeq 9.62$.

\subsection{The ultrarelativistic approximation}


In order to make quantitative statements about the critical
  solution beyond the energies covered by our self-force data, we
  shall assume that as $E\to\infty$, $\dot E$ diverges like the
  leading order of (\ref{Edotfit}), that is
\begin{equation}
\label{Edotultra}
{\dot E\over \eta}\simeq - 3^{\alpha-5}a_0(r-3)^{-\alpha}.
\end{equation}
(As discussed above, our qualitative conclusions that $\hat\varphi$
and $\hat\tau$ are finite at $E=\infty$ depend only on
$(r-3)^{-2}/\dot E$ and $(r-3)^{-3/2}/\dot E$ being integrable as
$r\to 3$.)  We shall refer to calculations where we only use the
leading power of $(r-3)$ throughout as the ultrarelativistic
approximation, and in this context we shall treat $a_0>0$ and
$\alpha>1$ as undetermined.

Combining the leading order of (\ref{dphidr}) with (\ref{Edotultra})
and integrating, we obtain
\begin{equation}
\label{phiofrapprox}
\hat\varphi(r)\simeq {3^{7/2-\alpha}\over
  2a_0(\alpha-1)}(r-3)^{\alpha-1},
\end{equation}
Similarly, taking the leading power of $(r-3)$ in (\ref{drdtau}), we
obtain
\begin{equation}
\label{tauofrapprox}
\hat\tau(r)={3^{9/2-\alpha}\over a_0(2\alpha-1)}(r-3)^{2\alpha-1\over 2}.
\end{equation}
In the ultrarelativistic approximation we can combine (\ref{Eultra})
with (\ref{phiofrapprox}) and (\ref{tauofrapprox}) to express $r-3$,
$\hat\tau$, $\hat\varphi$ and $E$ all as powers of each other. In particular
\begin{equation}
\label{Eoftauapprox}
E(\hat\tau)\simeq C_E \hat\tau^{-\beta}, 
\end{equation}
where 
\begin{equation}
C_E:=[3^{2\alpha-5}a_0(2\alpha-1)]^{-\beta},
\qquad \beta:={1\over 2\alpha-1}
\end{equation}
which will be used later. With
\begin{equation}
\dot\varphi\simeq {E\over\sqrt{3}}
\end{equation}
in the ultrarelativistic approximation, we also have
\begin{equation}
\label{varphioftauandE}
\hat\varphi\simeq {C_E\over\sqrt{3}(1-\beta)}\hat\tau^{1-\beta}
\simeq {C_E^{1/\beta}\over\sqrt{3}(1-\beta)}
E^{1-1/\beta}.
\end{equation}
In the ultrarelativistic limit (\ref{GammaV}) becomes
\begin{equation}
\label{Gammaultra}
\Gamma\simeq{9\over L}\simeq {\sqrt{3}\over E}
\simeq {\sqrt{3}\over C_E} \hat\tau^\beta, 
\end{equation}
where we have used (\ref{Eoftauapprox}) in the last equality.


\subsection{Validity of the self-force and adiabatic approximations}


The self-force approximation to the equations of motion is valid when
the total energy $mE$ of the small object is much smaller than the
total energy $M$ of the large black hole, that is, for
\begin{equation}
\label{selfforcevalid}
E\ll \eta^{-1}. 
\end{equation}
Because the self-force diverges as $E\to\infty$ at fixed $\eta$, we
must also estimate where the adiabatic approximation breaks down at
high energy. Heuristically, we assume that the adiabatic approximation
is valid if $E$ changes at most by some small fraction $\delta\ll 1$
per orbit, or
\begin{equation}
\label{Edotapprox}
-{d\ln E\over d \hat\varphi}\ll {\delta\over 2\pi\eta}.
\end{equation}
(The explicit $\eta$ on the right comes from the hat on the left.)
From (\ref{varphioftauandE}), we find that in the ultrarelativistic
regime this is equivalent to
\begin{equation}
\label{selfforceadiabatic}
E \ll\left({\eta\over\delta}\right)^{-{1\over 2\alpha-2}}.
\end{equation}
We see that the adiabaticity condition (\ref{selfforceadiabatic})
always implies the condition (\ref{selfforcevalid}) for the self-force
approximation itself to be valid if $\alpha\ge 3/2$. (Recall our
theoretically motivated value of $\alpha$ is $3/2$, and our fitted
value is $1.77$.) 

Note that the adiabatic approximation breaks down also as the plunge
is approached, because the effective potential becomes shallower,
leading to a rapid increase in $r_1$ at finite $F^r$ in (\ref{Fr1eqn})
\cite{OriThorne}.


\section{Critical scaling
in the self-force approximation}
\label{section:scaling}



\subsection{Perturbations of the critical solution}


We now consider a small perturbation of the critical solution that
results from imperfect fine-tuning of the initial data, as discussed
above in Sec.~\ref{section:dynsim}, namely
\begin{equation}
r=r_*(\hat\tau)+\delta r(\tau), \qquad L=L_*(\hat\tau)+\delta L(\tau),
\end{equation}
where $r_*$ and $L_*$ are given by the adiabatic ansatz
(\ref{rk},\ref{Lk}) above, while $\delta r$ and $\delta L$ are just
functions of $\tau$, that is, they are assumed to remain ``fast'' even
in the limit of weak radiation reaction. The self-acceleration is
perturbed as
\begin{equation}
F^a=F^a_*+O(\eta) O(\delta r,\delta L),
\end{equation}
where the factor $\eta$ arises simply because any self-acceleration scales as
$O(\eta)$. Substituting this into the equations of motion and
subtracting the critical solution, we formally obtain
\begin{eqnarray}
\delta \ddot r&=&-{1\over 2}[V_{,rr-}(L_0,r_0)\delta
  r+V_{,rL}(L_0,r_0)\delta L] \nonumber \\ &&+O(\delta r^2,\delta
L^2,\delta r\delta L)+O(\eta)O(\delta r,\delta L),\\
\delta \dot L &=& O(\eta)O(\delta r,\delta L).
\end{eqnarray}
We now neglect the error terms above, the first set as they are
nonlinear in $(\delta r,\delta L)$, the second set as they are small
corrections to the {\em coefficients} of a linear differential
equation for $(\delta r,\delta L)$.

In identifying the perturbed solution with the background critical
solution, we can adjust the origin of $\tau$ in the perturbed
solution. (This is a remnant of the usual gauge-dependence of
spacetime perturbations.) As $L_*(\hat\tau)$ is not constant, we can
use this to set $\delta L=0$ at any one time, and hence, by $\delta
\dot L\simeq 0$, at all times. Note that the neutral mode $\delta L=\rm
constant$ present in the geodesic approximation, which linked
neigbouring critical solutions, has become a gauge mode in the
presence of radiation reaction, where there is a single critical
solution, and that we have used a simple form of gauge fixing to
eliminate this mode. From now on we denote $\delta r$ in this gauge by
$x$. We are left with the single equation of motion
\begin{equation}
\label{ddotx}
\ddot x \simeq \Gamma(\hat\tau)^{-2}x,
\end{equation}
which is equivalent to (\ref{ddotxbis}) for the geodesic case, but
where now
\begin{equation}
\label{critGamma}
\Gamma(\hat\tau):=\Gamma[L_0(\hat\tau)]
\end{equation}
is time-dependent. 

\begin{figure}
\includegraphics[width=8cm]{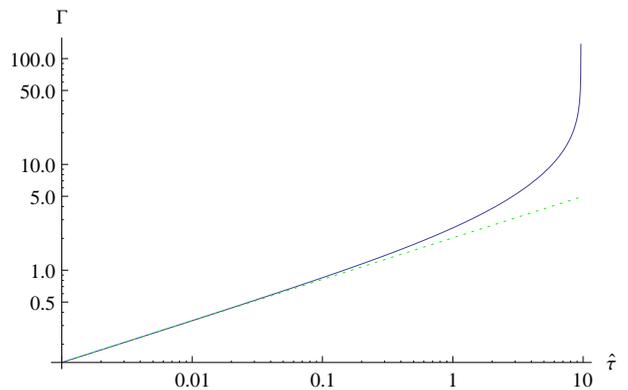}
\caption{Log-log plot of $\Gamma(\hat\tau)$ [defined in
    (\ref{GammaV}], based on (\ref{Edotfit},\ref{fitvals}) (solid
  line), and the ultrarelativistic approximation (\ref{Gammaultra})
  (dotted line). (The divergence of $\Gamma$ comes from the
  shallowness of the potential at $r=6$.)}
\label{figure:Gamma_vs_tauhat}
\end{figure}

\begin{figure}
\includegraphics[width=8cm]{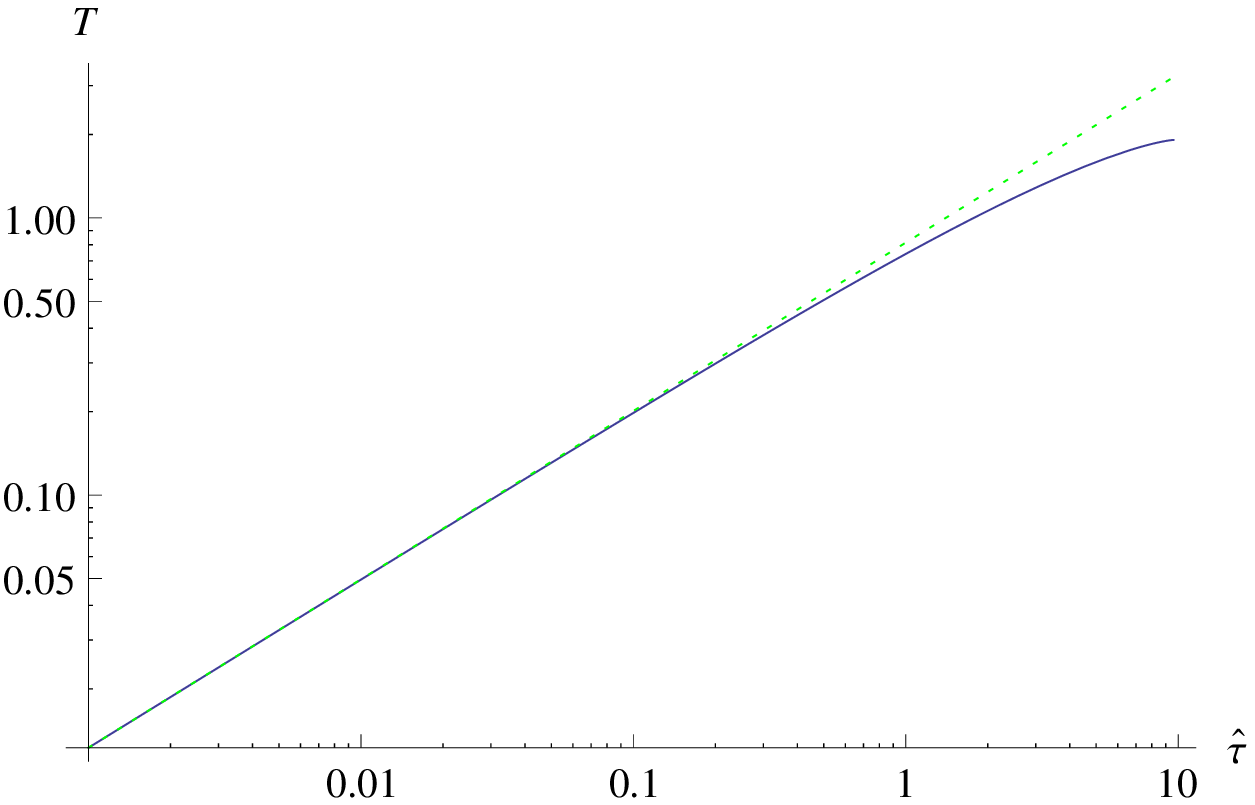}
\caption{Log-log plot of $T(\hat\tau)$ [defined in (\ref{Tdef}], based on
  (\ref{Edotfit},\ref{fitvals}) (solid line), and the ultrarelativistic
  approximation (\ref{Tultra}) (dotted line).}
\label{figure:That_vs_tauhat}
\end{figure}

Rewriting the equation entirely in terms of $\hat\tau$, we have
\begin{equation}
\label{xprimeprime}
x''\simeq \eta^{-2}\Gamma(\hat\tau)^{-2}\  x,
\end{equation}
where as before a prime denotes $d/d\hat\tau$. As $\eta$ is small, we
can use a WKB approximation. In terms of the ``WKB time''
\begin{equation}
\label{Tdef}
T(\hat\tau):=\int{d\hat\tau\over\Gamma(\hat\tau)}, 
\end{equation}
where for definiteness we let $T(0)=0$, the first-order WKB
approximation to the growing and decaying solutions is
\begin{equation}
\label{WKB} 
x_\pm(\hat\tau)\simeq \sqrt{\Gamma(\hat\tau)} e^{\pm \eta^{-1} T(\hat\tau)}.
\end{equation}
This is a good approximation for $|\dot\Gamma|\ll 1$, when the
prefactor varies much more slowly than the exponential. The functions
$\Gamma(\hat\tau)$ and $T(\hat\tau)$ are shown in
Figs.~\ref{figure:Gamma_vs_tauhat} and \ref{figure:That_vs_tauhat}.

In the ultrarelativistic approximation, from (\ref{Gammaultra}), we
have 
\begin{equation}
\label{Tultra}
T \simeq {C_E\over\sqrt{3}(1-\beta)}\hat\tau^{1-\beta}
\simeq {C_E^{1/\beta}\over \sqrt{3}(1-\beta)} E^{1-1/\beta}
\simeq\hat\varphi,
\end{equation}
where we have used (\ref{varphioftauandE}) in the last two equalities.
In the same approximation, 
\begin{equation}
\label{Gammadotapprox}
\dot\Gamma\simeq \sqrt{3}\beta C_E^{1-2\alpha}\eta E^{2\alpha-2}.
\end{equation}
Hence $|\dot\Gamma|\ll 1 $ for 
\begin{equation}
\label{WKBvalid}
E\ll \eta^{-{1\over 2\alpha-2}},
\end{equation}
which has the same power of $\eta$ as the criterion
(\ref{selfforceadiabatic}) for the critical solution to evolve
adiabatically. Clearly, with $\delta<1$, (\ref{selfforceadiabatic})
implies (\ref{WKBvalid}) and hence, for $\alpha\ge 3/2$, also
(\ref{selfforcevalid}). Hence the only criterion we need to impose is
(\ref{selfforceadiabatic}) (the critical solution evolves
adiabatically), and we can then always use the WKB approximation to
the critical solution.

For $\tau$ small enough ($E$ large enough) that $|\dot\Gamma|\gg 1$,
the WKB approximation does not hold. However, we are then necessarily
in the ultrarelativistic limit such that the approximation
(\ref{Gammaultra}) for $\Gamma(\hat\tau)$ holds, and with this,
(\ref{ddotx}) can be solved in closed form in terms of Bessel
functions. As this result is not directly relevant for our main
result, we give it in Appendix~\ref{appendix:xBessel}.


\subsection{Dependence on the initial data}
\label{section:initialdata}


For definiteness, we parameterise the space of initial data (modulo a
shift in time) by the initial energy at infinity $E_i$ and the impact
parameter $b$. We consider a generic near-critical solution in the
whirl phase $\tau_i<\tau<\tau_f$ where it is close to the critical
solution, so that
\begin{equation}
\label{deltaransatz}
x(\tau)\simeq A(E_i,b){x_-(\tau)\over x_-(\tau_i)}
+B(E_i,b){x_+(\tau)\over x_+(\tau_i)}.
\end{equation}
The factors $x_\pm(\tau_i)$ have been introduced for later convenience. 

$A$ and $B$ are smooth functions of $E_i$ and $b$, and by definition
$B$ vanishes for critical initial data. Therefore,
\begin{equation}
\label{approxAB}
A=A_0(E_i)+O(\delta b), \quad B=B_1(E_i)\delta b + O(\delta b)^2,
\end{equation}
where 
\begin{equation}
\delta b:=b-b_*(E_i,\eta).
\end{equation}
[The critical value $b_*(E_i,\eta)$ of the impact parameter is known
  in closed form only for $\eta=0$, when it is given by (\ref{b*E}),
  but its value does not matter in the following.]  We must have $A>0$
and $B_1>0$ because $x$ is positive and decreasing during the approach
to the critical solution, and then becomes negative in the capture
case $b<b_*$, or returns to large positive values in the scattering case
$b>b_*$.

Let the beginning and end of the whirl phase be defined by
$x(\tau_i)=X_i$ and $x(\tau_f)=X_f$, where $X_i$ and $X_f$ will be
characterised later.  If the whirl phase is sufficiently long, we can
neglect the growing mode at $\tau_i$ and the decaying mode at
$\tau_f$, so that
\begin{eqnarray}
\label{A0eqn}
A_0 &\simeq &X_i, \\
\label{B1eqn}
B_1\delta b &\simeq &X_f {x_+(\tau_i)\over x_+(\tau_f)}.
\end{eqnarray}
Note that $X_i$ is always positive, while $X_f$ is positive for
initial data that scatter and negative for initial data that plunge.

In the general discussion of critical scaling in
Sec.~\ref{section:dynsim}, we introduced two constants $C_1$ and $C_2$
that were treated as unknown, and combined into a constant $C_3$ in
the final result (\ref{C3def}). The equivalents of $C_1$ and $C_2$
in the current specific problem are $B_1$ and $X_f$,
respectively, and we shall now see that under certain conditions we
can determine both, so that there is no undetermined constant left in
our final results.

In order to determine $B_1$ we can use an energy argument if we
approximate $E$ as constant during the approach phase. This is
justified when the energy loss during the approach phase is small
compared to the energy loss during an extended whirl phase.  We
approximate (\ref{WKB}) over a short time interval (covering the
transition from approach to whirl) as
\begin{equation}
\label{approxexponential}
x_\pm(\tau)\simeq e^{\pm
      (\tau-\tau_i)/\Gamma(\tau_i)} x_\pm(\tau_i).
\end{equation}
Perturbing (\ref{Edef3}), which holds for geodesics, with respect to
$b$, we obtain
\begin{equation}
\label{ABresult}
E^2(L,b)-E_*^2(L)\simeq\dot x^2-\Gamma^{-2}x^2 \simeq -4\Gamma^{-2}AB,
\end{equation}
where the last equality follows from (\ref{approxexponential}). This
equation becomes exact in the geodesic limit, where $E$ is conserved
and (\ref{approxexponential}) is exact. On the other hand, when $E$ is
conserved, from (\ref{bLE}) we also have
\begin{equation}
\label{deltaEdeltab}
E^2(L,b)-E_*^2(L)\simeq-{2L^2\over b_*^3}\delta b 
\simeq -2[E_*^2(L)-1]{\delta b\over b_*}.
\end{equation}
Note the factor $E^2-1$ in $dE/db$. It comes about because geodesic
orbits with $E=1$ are at rest at infinity, and so the impact parameter
is not defined. More concretely, displacing a particle at rest at
infinity in a direction orthogonally to the line of sight to the large
black hole before dropping it corresponds to a rotation of the binary
system and makes no difference to the infall. Because of this factor,
some of our following results will be formally singular at $E=1$. This
is only because $b$ becomes a bad parameter of the initial data. In
practice, we are interested in initial data with $E>1$ (and in
particular $E\gg 1$), when this is not a problem.

We now approximate $E_*(L)=E_i$ in (\ref{deltaEdeltab}) and
$\Gamma\simeq \Gamma[L_*(E_i)]=:\Gamma_i$ in (\ref{ABresult}) because
we approximate $E$ and $L$ as constant for $\tau<\tau_i$ and because
$L_i\simeq L_*(E_i)$ by fine-tuning. Hence we find
\begin{equation}
\label{A0B1eqn}
A_0B_1\simeq {\Gamma_i^2(E_i^2-1)\over 2b_*}.
\end{equation}

From (\ref{A0eqn}), (\ref{B1eqn}) and (\ref{A0B1eqn}), and taking the
logarithm for later convenience, we now obtain
\begin{equation}
\label{scalingmaster}
-\ln{|\delta b|\over b_*}-\ln\left|{\Gamma_i^2(E_i^2-1)\over 2X_i
  X_f}\right| \simeq \ln{x_+(\tau_f)\over x_+(\tau_i)}. 
\end{equation}
This is our master equation for critical scaling: in principle, it
determines $E_f$ in terms of the mass ratio $\eta$ and the initial
data $(E_i,b)$. In order to obtain a definite result, however, we
still need to fix $X_i$ and $X_f$. As we discuss below, there is a
natural criterion for fixing them in the radiation reaction, but only
a more arbitrary one in the geodesic limit. Moreover, the criterion we
choose in the radiation reaction case has a singular limit as $\eta\to
0$ in the geodesic limit. We therefore discuss both cases separately,
starting with the geodesic case.


\subsection{Critical exponents in the geodesic case}
\label{section:geodesicscaling}


The critical solution is by definition as circular as possible in the
presence of radiation reaction, sitting approximately at the top of
the potential. Hence one possible criterion for $X_i$ and $X_f$ is
that the radial velocity be a given small fraction $C_e$ (for
``eccentricity'') of the tangential velocity, or
\begin{equation}
\label{Cedef}
\left({|\dot x|\over r\dot\varphi}\right)_{i,f}=C_e.
\end{equation}
Note that $C_e$ plays the role of $C_2$ in our general discussion in
Sec.~\ref{section:dynsim}, and while it must be small, its choice is
arbitrary -- we have introduced it here because it is more
intuitive than $X_{i,f}$. We have
\begin{equation}
\label{Cevalue}
\left({|\dot x|\over r\dot\varphi}\right)_{i,f}\simeq{\sqrt{E^2-V}\over 
    r^{-1} L}
\simeq {|X_{i,f}|\over\Gamma r^{-1} L}
\simeq {|X_{i,f}|\over\sqrt{3}\Gamma E},
\end{equation}
where we have used (\ref{Ldefbis}) and (\ref{Edef3}) in the first
step, (\ref{GammaV}) in the second step, and the third approximation
holds only in the ultrarelativistic limit.

In the geodesic case, where the background solution is time
independent,
\begin{equation}
x_\pm=e^{\pm \tau/\Gamma}
\end{equation}
holds exactly. Using this and (\ref{scalingmaster}-\ref{Cevalue}), we
obtain
\begin{equation}
\label{geodesictauscaling}
\tau_f-\tau_i\simeq -\Gamma \ln{|\delta b|\over b_*}
-\Gamma \ln \tilde C_e.
\end{equation}
were $\Gamma$ is given by (\ref{GammaV}). Here
\begin{equation}
\tilde C_e:={E^2-1\over 2C_e^2 r^{-2}L^2}
={4-r\over 2C_e^2 r}\simeq {1\over 6C_e^2},
\end{equation}
where the last approximate equality holds only in the
ultrarelativistic limit $r\simeq 3$. Clearly, $\tilde C_e$ inherits the
arbitrariness of $C_e$. The result (\ref{geodesictauscaling}) can be
expressed in terms of orbits as
\begin{equation}
\label{geodesicnscaling}
n\simeq -\gamma\ln{|\delta b|\over b_*}
-\gamma \ln \tilde C_e,
\end{equation}
where 
\begin{equation}
\label{gammageodesic}
\gamma := {\dot\varphi\over 2\pi}\Gamma={ L^{1/2}\over 2\pi
  L_{12}^{1/2}}
\end{equation}
In the ultrarelativistic limit, we have
\begin{eqnarray}
\Gamma &\simeq& {\sqrt{3}\over E}, \\
\label{gammaultra}
\gamma &\simeq& {1\over 2\pi}.
\end{eqnarray}

The critical exponents $\Gamma$ and $\gamma$ for the geodesic limit
were previously obtained by Pretorius and Khurana
\cite{PretoriusKhurana} for equatorial geodesics of Kerr, and had
previously been given, in different notation, in Eq.~(2.25) of
  \cite{CutlerKennefickPoisson}. The additive constant parameterised
  by $C_e$ could only be determined more exactly by a full modelling
  of the perturbed orbit to replace the artifical split into a
  zoom-in, whirl and zoom-out or plunge phase.

As an extension of the result for the time spent whirling, we can --
trivially -- estimate the energy loss during the whirl phase as
\begin{equation}
\label{geodesicEscaling}
\Delta E\simeq (\tau_f-\tau_i) \dot E(E_i) ,
\end{equation}
where $(\tau_f-\tau_i)$ is given in terms of the initial data by
(\ref{geodesictauscaling}) and the constant $\dot E(E_i)$ is given
implicitly by relating $E_i$ to $r$ using the exact formula (\ref{Ec})
for the energy on a circular orbit and $\dot E$ to $r$ using our
numerical result (\ref{Edotapprox}) for the energy loss on circular
geodesics. This approximation, where we neglect the radiation reaction
on the orbit itself, is therefore consistent only for $\Delta E\ll
E_i$. (\ref{geodesicEscaling}) complements the results given in
\cite{Bertietal2010} for the energy radiated in the geodesic
approximation scattering orbits not close to the threshold of
immediate merger.


\subsection{Generalised critical scaling in the radiation reaction case}
\label{section:results}


With radiation reaction, a less arbitrary definition of $X_i$ and
$X_f$ is given by equating the rate of change of the perturbation $x$
with that of the background solution, that is
\begin{equation}
\label{xixfdef}
|\dot x|\simeq |\Gamma_{i,f} X_{i,f}|\simeq \dot r_*. 
\end{equation}
Furthermore, we have seen above that we can always use the WKB
approximation (\ref{WKB}) to the perturbation modes when the adiabatic
self-force approximation to the background critical solution is valid.
Substituting our new definition (\ref{xixfdef}) of $X_{i,f}$ into the
left-hand side of the scaling master equation (\ref{scalingmaster}),
and the WKB approximation (\ref{WKB}) into its right-hand side,
(\ref{scalingmaster}) becomes
\begin{equation}
\label{scalingmasterbis}
-\ln{|\delta b|\over b_*}-\ln {\Gamma_i^3\Gamma_f(E_i^2-1)\over 2
  \eta^{-2} r_{0i}'r_{0f}'} \simeq 
{T_f-T_i\over \eta}
+{1\over 2}\ln{\Gamma_f\over\Gamma_i}.
\end{equation}
We can re-order this as
\begin{equation}
\label{scalingbyeta}
-\ln{|\delta b|\over b_*} \simeq {T_f-T_i\over \eta}
+2\ln\eta + K_i + K_f ,
\end{equation}
where we have defined the shorthands
\begin{eqnarray}
\label{Kidef}
K_i&:=& \ln {\Gamma_i^{5/2}(E_i^2-1)\over 2 r_{0i}'}, \\
\label{Kfdef}
K_f&:=& \ln {\Gamma_f^{3/2}\over r_{0f}'}.
\end{eqnarray}
Here we consider $T$, $K_i$ and $K_f$ as functions of $\hat\tau$ (or
any of the other slow variables $E$, $r$, $\hat\varphi$). The only
$\eta$-dependence is then the one explicitly shown in (\ref{scalingbyeta}). 

Alternatively, we can re-order (\ref{scalingbyeta}) as
\begin{equation}
\label{scalingbyif}
{T_f\over \eta} + K_f \simeq -\ln{|\delta b|\over b_*} + {\hat
  T_i\over \eta} - K_i - 2\ln\eta.
\end{equation}
This gives a known function of $\hat\tau_f$ (or $E_f$), on
the left-hand side, as a known function of the mass ratio $\eta$ and the
initial data $(E_i,b_i)$, on the right-hand side. To fully expand this
in terms of $E_f$ and $E_i$, we first note that (\ref{Ec}) gives an
expression for $E(r)$ that can be inverted to obtain
$r(E)$. (\ref{GammaV}) gives $\Gamma(L)$ and (\ref{Lc}) gives $L(r)$,
so that we can explicitly obtain $\Gamma(E)$. (We do not give the
expression here, as it is unwieldy.) Next we note that (\ref{drdtau})
can be rewritten as
\begin{equation}
r_0'(r)=[\eta^{-1}\dot E](r) \left({dE\over dr}\right)^{-1},
\end{equation}
where $r_0':=dr_0/d\hat\tau$ as before. Finally we note that the
definition (\ref{Tdef}) of $T(\hat\tau)$ can be rewritten as
\begin{equation}
\label{Tofr}
T(r)=\int{dr\over r_0'(r) \Gamma(r)},
\end{equation}
which can then be written as $T(r(E))$, or directly as
\begin{equation}
\label{TofE}
T(E)=\int{dE\over r_0'(r(E)) \Gamma(r(E)) {dE\over dr}(r(E))}.
\end{equation}
With the definitions (\ref{Kidef},\ref{Kfdef}) of $K_i$ and $K_e$ in
terms of $\Gamma$, $E$ and $r_0'$, we can now write our master
equation (\ref{scalingbyif}) in terms of the mass ratio $\eta$, the
initial data $(E_i,b)$, and the final energy $E_f$, subject to
evaluating the integral in either (\ref{Tofr}) or
(\ref{TofE}). However, this equation cannot be solved for
$E_f=E_f(\eta,E_i,b)$ in closed form.

\begin{figure}
\includegraphics[width=8cm]{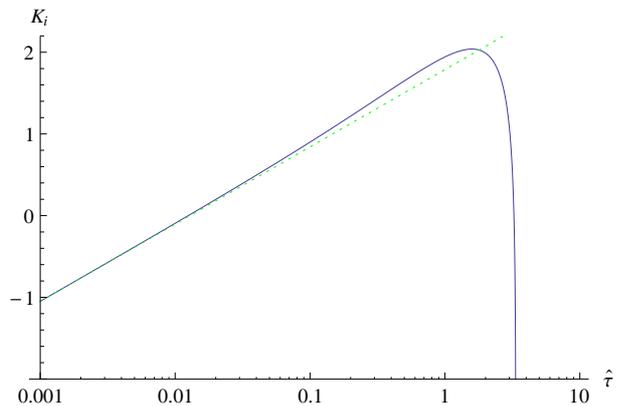}
\caption{Log-linear plot of $K_i(\hat\tau)$ [defined in (\ref{Kidef}],
  based on (\ref{Edotfit},\ref{fitvals}) (solid line), and the
  ultrarelativistic approximation (\ref{Kiultra}) (dotted line). The
  full curve ends because of the factor $E^2-1$ in the exact
  expression, which is approximated as $E^2$ in the ultrarelativistic
  expression}
\label{figure:Ki_vs_tauhat}
\end{figure}

\begin{figure}
\includegraphics[width=8cm]{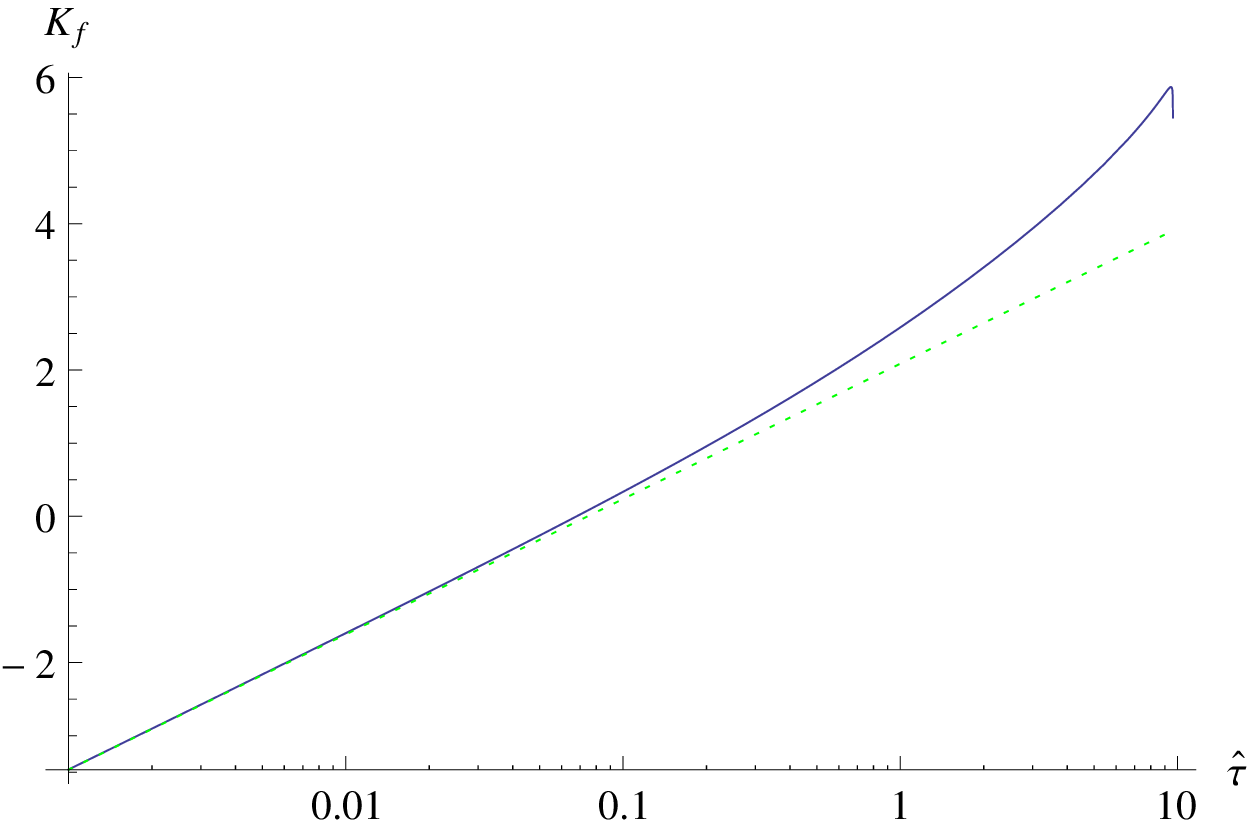}
\caption{Log-linear plot of $K_f(\hat\tau)$ [defined in (\ref{Kfdef}],
  based on
  (\ref{Edotfit},\ref{fitvals}) (solid line), and the ultrarelativistic
  approximation (\ref{Kfultra}) (dotted line).}
\label{figure:Kf_vs_tauhat}
\end{figure}

\begin{figure}
\includegraphics[width=8cm]{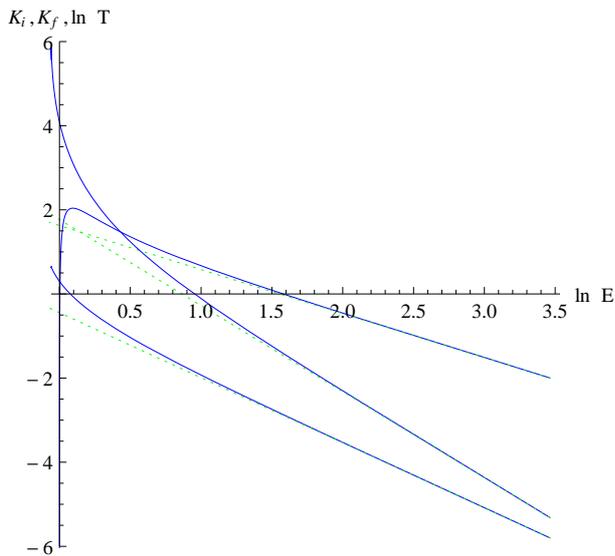}
\caption{Plot of (from top to bottom at large $E$) $K_i$, $K_f$ and
  $\ln T$ (note the logarithm), all versus $\ln E$, based on
  (\ref{Edotfit},\ref{fitvals}) (solid lines), and the corresponding
  ultrarelativistic approximations (dotted lines).}
\label{figure:all_vs_lnE}
\end{figure}

We can make (\ref{scalingbyif}) more explicit, in terms of only
powers and logarithms, when the entire whirl phase is
ultrarelativistic, that is $E_f\gg 1$. We have already given the 
ultrarelativistic approximations for $T(\hat\tau)$, $\hat
T(E)$ and $T(\hat\varphi)$ in (\ref{Tultra}), and we similarly
find that that in the ultrarelativistic limit
\begin{eqnarray}
\label{Kiultra}
K_i&\simeq& -2\ln 2+\left({3\over4}+6\beta\right)\ln 3 \nonumber \\ &&
-{3\beta\over 2}\ln a_0
+\left(1-{3\beta\over 2}\right)\ln{\hat\tau\over\beta}, \\
\label{Kfultra}
K_f&\simeq& -\ln 2+\left({5\over4}+2\beta\right)\ln 3 \nonumber \\ &&
-{\beta\over 2}\ln a_0
+\left(1-{\beta\over 2}\right)\ln{\hat\tau\over\beta}.
\end{eqnarray}
[These can be readily re-expressed in terms of $\hat\varphi$, $E$, or
  $(r-3)$ using (\ref{phiofrapprox}-\ref{Eoftauapprox}).]  We have
plotted $T(\hat\tau)$ above in Fig.~\ref{figure:That_vs_tauhat},
and we plot $K_i(\hat\tau)$ and $K_f(\hat\tau)$ in
Figs.~\ref{figure:Ki_vs_tauhat} and \ref{figure:Kf_vs_tauhat}, all
based on (\ref{Edotfit},\ref{fitvals}) and each compared against its
ultrarelativistic approximation. Fig.~\ref{figure:all_vs_lnE} plots
all three against $E$ instead of $\hat\tau$.

Note from Fig.~\ref{figure:all_vs_lnE} that $\ln T$, $K_i$ and
$K_f$ vary comparably rapidly along the critical solution, but it is
$T/\eta$ (without the logarithm), $K_i$ and $K_f$ that appear in
our scaling result (\ref{scalingbyeta}). Hence one can suppress $K_i$
and $K_f$ in (\ref{scalingbyeta}) as logarithmic corrections relative
to the $T$ terms. Doing this, suppressing also the constant term
$2\ln\eta$, and taking the ultrarelativistic limit for clarity
reduces (\ref{scalingbyeta}) to
\begin{eqnarray}
-\ln{|\delta b|\over b_*}
&\simeq&{T_f-T_i\over \eta}  \\
\label{tauscalingultra}
&\simeq&\eta^{-\beta}{C_E\over\sqrt{3}(1-\beta)}
\left(\tau_f^{1-\beta}-\tau_i^{1-\beta}\right) \\
\label{Escalingultra} 
&\simeq&\eta^{-1}{C_E^{1/\beta}\over\sqrt{3}(1-\beta)}
\left(E_f^{1-1/\beta}-E_i^{1-1/\beta}\right) \\
\label{varphiscalingultra}
&\simeq&\varphi_f-\varphi_i. 
\end{eqnarray}
It is interesting to note that the amount of fine-tuning of the
initial data (measured by $\ln|\delta b|$) required to achieve a given
number of orbits on the critical solution is independent (within the
self-force approximation) of the mass-ratio $\eta$. [In particular,
  the limit $\eta\to 0$ of (\ref{varphiscalingultra}) is trivial and
  gives (\ref{geodesicnscaling}) with (\ref{gammaultra}).] By
comparison, better fine-tuning (smaller $|\delta b|$) is required to
achieve a given time on the critical solution, or a given amount of
energy radiated, as $\eta$ becomes smaller. These scalings with $\eta$
are not immediately intuitive.

Finally, to see how the self-force result is related to the geodesic
limit, we go back to (\ref{scalingmasterbis}). The second term on the
left-hand side comes from the criterion (\ref{xixfdef}) for
determining $X_i$ and $X_f$, and is simply not defined when the
self-force is neglected. (Formally, it diverges as $\eta\to 0$.)  In
the geodesic case we therefore adopted the more ad-hoc definition
(\ref{Cedef}) of $X_i$ and $X_f$. Hence, in the geodesic limit the
left-hand side of (\ref{scalingmasterbis}) becomes $-\ln(|\delta
b|/b_*)$ plus some constant related to $X_i$ and $X_f$ but independent
of $\delta b$. On the right-hand side of (\ref{scalingmasterbis}), we
note $\Gamma_i=\Gamma_f=:\Gamma$, so the second term vanishes, and
from (\ref{Tdef}) we have
\begin{equation}
\lim_{\eta\to 0}{T_f-T_i\over
  \eta}={\tau_f-\tau_i\over\Gamma}.
\end{equation}
Hence we recover (\ref{geodesictauscaling}) as the $\eta\to 0$ limit
of (\ref{scalingmasterbis}), but with the constant $\tilde C_e$
undetermined by (\ref{scalingmasterbis}). 


\section{Conclusions}
\label{section:conclusions}

Pretorius and Khurana \cite{PretoriusKhurana} suggested equatorial
zoom-whirl orbits in Kerr as a toy model for the critical phenomena at
the threshold of immediate merger in equal mass black hole binaries
that they had observed in numerical relativity simulations. Here, we have
adapted this toy model to make {\em quantitative} predictions in the
limit where one of the two black holes is much larger than the
other. We can then model the smaller black hole as a point particle of
mass $m$ in a background spacetime given by the larger one (assumed to
be Schwarzschild with mass $M=1$), and moving on an orbit determined
by a gravitational self-force proportional, to leading order, to the
mass ratio $\eta:=m/M\ll 1$.

In our model there is a single critical solution that evolves
adiabatically along the sequence of unstable circular geodesics,
beginning with infinite energy and ending with a plunge when almost
all the energy has been dissipated in gravitational waves. In this
process, the critical orbit evolves from $r=3$ and $E=\infty$ (the
light ring) to $r=6$ and $E=\sqrt{8/9}$ (the last stable circular
orbit), making only a finite number (approximately $0.41/\eta$) of
orbits.

Any initial data fine-tuned to the critical impact parameter are
attracted to the critical solution, joining it at the appropriate
energy. They leave it after a number of orbits determined by the
degree of fine-tuning, to either plunge or scatter. Beyond a certain
degree of fine-tuning, they stay on the critical solution until it
plunges. 

The type of critical solution we have described here is unfamiliar in
general relativity, being neither self-similar (type II) nor
stationary (type I). Rather, it avoids immediate plunge or scatter by
remaining as circular as is compatible with radiation reaction. The
appropriate symmetry ansatz is therefore a formal slow time expansion,
in which the orbit evolves adiabatically from one circular geodesic
orbit to an adjacent one under the effect of energy and angular
momentum loss. We might call this type Ia, for ``adiabatic''.

The fact that a type Ia critical solution evolves because of
dissipation makes no essential difference to the dynamical systems
picture, where the critical solution is defined as balanced between
two basins of attraction. The conjecture in \cite{HealyLevinShoemaker}
that ``zoom-whirl behaviour has better odds at surviving dissipation
[for smaller $\eta$] because of the slower dissipation time'' is not
supported by this picture, and in fact Eq.~(\ref{varphiscalingultra})
shows that the dependence of the whirl angle on the fine-tuning of the
initial data is independent of $\eta$ in the ultrarelativistic
limit.

Our main result links the final energy (per rest mass) $E_f$ of the
small object to its initial energy $E_i$, the impact parameter $b$ and
the mass ratio $\eta$. This is given in implicit form in our master equation
(\ref{scalingbyif}) -- see also the discussion in the paragraph
following this equation. We have also given more explicit approximate
expressions (neglecting subleading terms, and assuming for clarity
that $E_f\gg 1$) in Eqs.~(\ref{tauscalingultra}-\ref{varphiscalingultra}). 

For the purpose of our calculation, there are four different regimes
for the initial energy:

1) For $\sqrt{8/9}<E_i\lesssim 4$ we have accurate self-force data and
so our results are reliable, but cannot be expressed in closed form.
(We plot them, and the interested reader could reconstruct all our
plots from the data and formulas given in this paper.) Even in this
regime, we had to extend previous self-force calculations much closer
to the light ring at $r=3M$, and it is clear that the self-force
diverges there. It would be highly interesting in its own right to
understand the origin of this divergence rigorously. 

2) For $4\lesssim E_i\lesssim\eta^{-1/(2\alpha-2)}$, we are
extrapolating our numerical self-force data for $\dot E$ as
$(r-3)^{-\alpha}$, see Eq.~(\ref{Edotultra}). (A best fit to our
numerical data gives $\alpha\simeq 1.77$, but we also have a heuristic
theoretical argument for $\alpha=3/2$, which is a less good fit but
not obviously wrong, see Fig.~\ref{figure:dEdtau}.) In this regime the
motion is ultrarelativistic, which allows us to give explicit
expressions. As $\eta$ can be arbitrarily small, we give our
(extrapolated) results up to $E_i=\infty$.

3) If $\alpha<3/2$, there is an the even higher energy regime
$\eta^{-1/(2\alpha-2)}\lesssim E_i\lesssim \eta^{-1}$ when the self-force
approximation is still valid, but the adiabatic approximation is
not. (This regime is empty for $\alpha\ge 3/2$.) It should be
possible to calculate in this regime in the foreseeable future, but at
the moment of writing this, the gravitational self-force on
non-geodesic orbits in the Schwarzschild spacetime has not yet been
calculated. (Similarly, it should become possible in the foreseable
future to extend our self-force results to the case where the large
black hole is Kerr.)

4) Finally, for $E_i\gtrsim \eta^{-1}$, the (total, mostly kinetic)
energy of the small particle becomes comparable to that of the large
black hole, and the self-force approximation itself breaks down. (We
come back to this regime just below.)

Two surprising features of the critical solution at the threshold of
immediate merger, at least in the extreme-mass ratio approximation,
are its unfamiliar (approximate) symmetry, and the fact that the {\em
  fraction} of $E$ radiated per orbit increases with $E$ fast enough
that the critical solution evolves from $E=\infty$ (or at least from
$E\sim \eta^{-1}\gg 1$, for any fixed finite mass ratio $\eta$) to
plunge at $E\simeq\sqrt{8/9}$ in a finite number of orbits.

We conjecture that both these qualitative features hold also in the
comparable to equal mass ratio case. Evidence that the fraction of
energy radiated per orbit increases with energy is given by comparing
the numerical relativity results of \cite{PretoriusKhurana}
($1.0-1.5\%$ at a boost of $k\simeq 0.22$) and \cite{Sperhake} ($\sim
13\%$ at a boost of $k\simeq 1.5$ \cite{footnote}). We should stress
again that for any given mass ratio $\eta$ (and spins) there is only
one critical solution, which begins with infinite energy. We also
conjecture that this solution radiates an infinite amount of energy over a
finite, probably quite small, number of orbits. At low energies, an
adiabatic approximation should be possible, where the critical
solution evolves along a sequence of stationary solutions with a
helical Killing vector under the influence of radiation reaction, but
if our conjecture is correct this breaks down at sufficiently large
energy.

In the comparable mass case, the dependence of the orbital radius of
the critical solution on the energy will be very different from that
in the extreme mass ratio case. For two point masses $M_1:=M$ and
$M_2:=\eta M$ in special relativity, where we set $\eta\le 1$ by
definition, with a relative boost $k:=-u^a_1u_{2a}\ge 1$ towards each
other, the total energy in their common centre of mass frame is given
by
\begin{equation}
E_{\rm CM}=M\sqrt{1+2\eta k+\eta^2}.
\end{equation}
Furthermore, the hoop conjecture suggests that the critical impact
parameter for immediate merger is 
\begin{equation}
\label{bsimE}
b_*\sim E_{{\rm CM},i}, 
\end{equation}
and this is borne out by numerical relativity simulations for
comparable masses and large $k$ \cite{PretoriusChoptuik}. Finally, as
the critical solution is by definition always between scattering and
merger, we also expect 
\begin{equation}
\label{rsimE}
r_*(t)\sim E_{\rm CM}(t)
\end{equation}
along the critical solution. In the limit $k\eta\gg 1$, (\ref{bsimE})
gives us $b_*\sim M\sqrt{2\eta k}$ and from (\ref{rsimE}) the
critical solution {\em spirals in} from infinite orbital radius to the
innermost stable circular orbit (ISCO). In the opposite limit
$k\eta\ll 1$, (\ref{bsimE}) gives us $b_*\sim M$ (a more precise
result is of course $b_*\simeq 3\sqrt{3}M$), the self-force
approximation we have given here becomes valid, and the critical
solution {\em spirals out} from the light ring to the ISCO. 

Constructing the critical solution presents a challenge at any
mass-ratio: for self-force methods in the extreme mass-ratio case to
reach higher energies and to go beyond the adiabatic approximation to
the orbit, and for effective one-body and numerical relativity methods
in the comparable mass ratio case to construct the critical solution
even at low energies.


\acknowledgments


We thank Nori Sago for using his code to test some of our self-force data.
CG, SA and LB acknowledge support from STFC through grant numbers
PP/E001025/1 and ST/J00135X/1.


\appendix


\section{Numerical data}
\label{appendix:data}


We present here the numerical data used to inform our ``empirical''
formula (\ref{Edotfit}) for $\dot E(r)$. The data were generated using
the self-force code developed by Akcay in
Ref.~\cite{Akcay:2010dx}. The code computes $\dot E$ in two different
ways: (i) Directly, by calculating the component $F_t$ of the local
gravitational self-force (per unit particle mass $m$) and using $\dot
E=-F_t$; and indirectly, by numerically evaluating the flux of energy
in gravitational waves radiated to infinity and down the event
horizon, and using the energy balance relation $-\dot E = (dE_{\rm
  GW}/dt) (dt/d\tau)$. Here $dE_{\rm GW}/dt$ is the total flux of
energy (per $m$) carried by the gravitational waves, and
$dt/d\tau=(1-3/ r)^{-1/2}$. $dE_{\rm GW}/dt$ is made up of two pieces,
$dE^+_{\rm GW}/dt$ and $dE^{-}_{\rm GW}/dt$, respectively associated
with radiation going out to infinity and absorbed by the black
hole. The numerical evaluation of both $F_t$ and $dE^{\pm}_{\rm
  GW}/dt$ are detailed in Ref.~\cite{Akcay:2010dx}. We find that the
two computations agree extremely well, differing by no more than one
part in $\sim 10^{10}$ for $r\geq 3.2$. At smaller radii the agreement
is less good, but for none of the radii we considered (down to
$r=3.02$) is the disagreement greater than $1\%$.

There is a practical limit on how close to $r=3$ our numerical method
can reach. Both above computations (of the local self-force and of the
asymptotic fluxes) rely on a summation over multipole-mode
contributions, with a practical cut-off at some $l=l_{\rm max}$. Our
calculations typically use $l_{\rm max}\simeq 80$ for the self-force and
$l_{\rm max}\simeq 140$ for the fluxes. As the (ultra-relativistic)
limit $r\to 3$ is approached, a beaming-like effect broadens the
$l$-mode power distribution and shifts it toward large $l$. For $r-3$
small enough it becomes computationally prohibitive to obtain a
sufficient number of modes. In practice, we have not attempted to
obtain accurate data below $r\simeq 3.02$.

Table~\ref{table:data} displays our numerical results for a sample of
orbital radii in the range $3.02\leq r\leq 6$. We choose to show
$\dot{E}$ data from flux calculations, which, we believe, are slightly
more accurate. For completeness we also show the flux $dE^{-}_{\rm
  GW}/dt$ absorbed by the black hole as a fraction of the total flux
radiated. For circular orbits at the ISCO only very little ($\simeq
0.3\%$) of the energy emitted goes down the hole. This fraction,
however, becomes very significant for orbits closer to the light-ring,
reaching as much as $37.7\%$ at $r=3.02$. A rough extrapolation gives
$\simeq 42\%$ at the light ring itself.

\begin{table}[htb] 
\begin{tabular}{|l |l| c|}
\hline\hline
$r$ &  $-\eta^{-1}\dot E$ &    \% absorbed  \\
\hline\hline
3.02 & $ 83(1)  $ & $ 37.7  $  \\
3.03 & $ 39.9(2) $ & $ 36.3   $  \\
3.04 & $ 23.62(5)  $ & $ 35.1   $  \\
3.05 & $ 15.64(1)   $ & $ 34.1   $  \\
3.10 & $ 4.232549(9)  $ & $ 30.0  $  \\
3.15 & $ 1.9223864(1)   $ & $ 26.9    $  \\
3.20 & $ 1.081977467(7)   $ & $ 24.2  $  \\
3.30 & $ 0.467163243(4)   $ & $ 19.9  $  \\
3.40 & $ 0.25012891219(4)   $ & $ 16.3  $  \\
3.50 & $ 0.1510176014(1)   $ & $ 13.4  $  \\
3.60 & $ 0.09864553444(1) $ & $ 11.0   $  \\
3.70 & $ 0.06818392468(3) $ & $ 9.09  $  \\
3.80 & $ 0.0491937318(2)  $ &  $ 7.50   $  \\
3.90 & $ 0.03670796106(3) $ & $ 6.20   $  \\
4.00 & $ 0.02814331203(5)  $ & $ 5.15   $  \\
4.10 & $ 0.02206146488(5)  $ & $ 4.30  $  \\
4.20 & $ 0.01761664969(8)  $ &  $ 3.61  $  \\
4.30 & $ 0.01428856144(8)  $ &  $ 3.04   $  \\
4.40 & $ 0.01174466751(8)  $ &  $ 2.57   $  \\

4.50 & $ 0.00976536837(3)  $ &  $ 2.19   $  \\
4.60 & $ 0.00820146283(7)  $ &  $ 1.88   $  \\
4.70 & $ 0.00694902541(6)  $ &  $ 1.61   $  \\
4.80 & $ 0.00593406093(6)  $ &  $ 1.39   $  \\
4.90 & $ 0.00510284885(5)  $ &  $ 1.21   $  \\
5.00 & $ 0.00441570494(4)  $ &  $ 1.05   $  \\
5.10 & $ 0.00384285426(2)  $ & $ 0.922   $  \\
5.20 & $ 0.00336164181(3)  $ & $ 0.810   $  \\
5.50 & $ 0.00231155387(2)  $ & $ 0.562   $  \\
5.75 & $ 0.00173630600(1)  $ & $ 0.424   $  \\
6.00 & $ 0.00132984067(1)  $ & $ 0.326   $  \\ 
\hline\hline
\end{tabular}
\caption{Numerical data for the rate of energy loss, $\dot
  E=dE/d\tau$, for a sample of sub-ISCO circular geodesic orbits of
  radius $r$. In the second column parenthetical figures show the
  estimated uncertainty in the last displayed decimals. The third
  column shows the percentage of energy radiated down the event
  horizon, $100\times (dE^{-}_{\rm GW}/dt)/(dE_{\rm GW}/dt)$; all
  figures shown are significant.}
\label{table:data}
\end{table}


\section{Perturbation modes of the critical solution in the
  ultrarelativistic limit}
\label{appendix:xBessel}

\begin{figure}
\includegraphics[width=8cm]{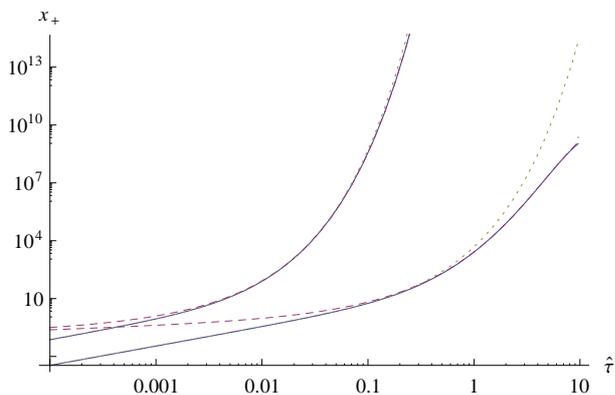}
\caption{Log-log plot of the growing perturbation mode $x_+(\hat\tau)$
  for $\eta=0.1$ (lower curves) and $\eta=0.01$ (upper curves). The
  solid lines are numerical solutions started up from
  (\ref{Besselxplus}) at small $\tau=0.0001$, the dashed lines are
  given by the WKB approximation (\ref{WKB}) with the {\em exact}
  values of $\Gamma(\hat\tau)$ and $T(\hat\tau)$, and the dotted
  lines are the ultrarelativistic (Bessel function) approximation
  (\ref{Besselxplus}), normalised to agree with the WKB
  approximation. The power-law growth of both the exact mode and its
  ultrarelativistic approximation at small $\tau$ visible here is in
  fact linear, as can be shown from (\ref{Besselxplus}). It can be
  seen that the WKB approximation is good at large $\hat\tau$/small
  $E$, and the ultrelativistic approximation in the other regime. Note
  that our self-force data are valid only for $\hat\tau\gtrsim 0.2$,
  corresponding to $r>3.02$. It is clear from the figure that starting
  the numerical integration with data from the ultrarelativistic
  approximation at $\hat\tau=0.2$ instead of $0.001$ would make little
  difference for $\eta=0.1$ and very little difference for
  $\eta=0.01$.}
\label{figure:xplus1and2}
\end{figure}
\begin{figure}
\includegraphics[width=8cm]{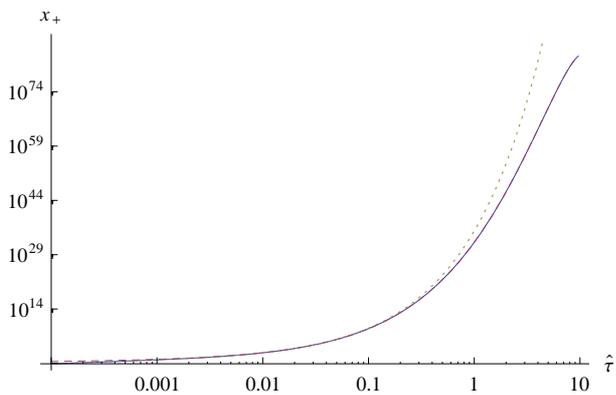}
\caption{Log-log plot of the growing perturbation mode $x_+(\hat\tau)$
  for $\eta=0.01$, showing the full range. The lines are as in
  Fig.~\ref{figure:xplus1and2}. From this and the preceding figure it
  is clear that the WKB approximation (\ref{WKB}) is excellent for
  $\hat\tau\gtrsim 0.2$ for $\eta=0.1$, and $\hat\tau\gtrsim 0.001$
  for $\eta=0.01$. Continuing this trend, the WKB approximation is
  good from smaller $\hat\tau$ (higher energies) for smaller $\eta$,
  and it is always good essentially up to plunge.}
\label{figure:xplus2}
\end{figure}

In the independent variable $T:=\eta^{-1} T$, with $\hat\tau$ defined
by (\ref{Tdef}), (\ref{ddotx}) becomes
\begin{equation}
\label{xTODE}
{d^2x\over dT^2}-{d\ln\Gamma\over dT}{dx\over
  dT}-x=0.
\end{equation}
The ultrarelativistic approximations (\ref{Gammaultra}) and
(\ref{Tultra}) give $\Gamma\propto T^{\beta/(1-\beta)}$, and
so in the ultrarelativistic approximation (\ref{xTODE}) becomes
\begin{equation}
{d^2x\over dT^2}-{\beta\over 1-\beta}{1\over T}{dx\over
  dT}-x=0.
\end{equation}
With the substitution
\begin{equation}
x(T)=:T^\nu y(T)\propto \hat\tau^{1/2} y(T), \qquad \nu:= {1\over 2(1-\beta)},
\end{equation}
we obtain
\begin{equation}
T^2{d^2y\over dT^2}+T{dy\over dT}-(T^2+\nu^2)y=0,
\end{equation}
which is the modified Bessel equation with index $\nu$. Hence two
linearly independent perturbation modes of the critical solution are
\begin{eqnarray}
\label{Besselxplus}
x_+(\hat\tau)&:=&\sqrt{2\pi\over \eta(1-\beta)}\hat\tau^{1/2}
I_\nu\left[\eta^{-1}T(\hat\tau)\right], \\
\label{Besselxminus}
x_-(\hat\tau)&:=&\sqrt{2\over\pi\eta(1-\beta)}\hat\tau^{1/2} 
K_\nu\left[\eta^{-1}T(\hat\tau)\right],
\end{eqnarray}
with $T(\hat\tau)$ in these formulas now {\em defined} as the
ultrarelativistic expression (\ref{Tultra}). The normalisations have
been chosen so that for large $\tau$ such that $|\dot\Gamma|\ll 1$,
these solutions have the asymptotic form (\ref{WKB}), with
$T(\hat\tau)$ defined by (\ref{Tultra}) and $\Gamma(\hat\tau)$ defined
by the ultrarelativistic expression (\ref{Gammaultra}). Hence we
recover the WKB approximation (\ref{WKB}) in the ultrarelativistic
limit. On the other hand, as $\tau\to 0$, $x_+\sim \hat\tau$ and
$x_-\sim 1$. Fig.~\ref{figure:xplus1and2} and \ref{figure:xplus2}
show that an excellent approximation to the true growing mode is given
by (\ref{Besselxplus}) for small $\tau$ and by (\ref{WKB}) for large
$\tau$, with the two approximations overlapping.

In numerically evolving $x_+$ towards increasing $\hat\tau$ from
initial data given at small $\tau$ by (\ref{Besselxplus}), we stably
find the growing WKB solution at large $\tau$. Finding $x_-$ over its
entire range is more difficult: in going forward from
(\ref{Besselxminus}), numerical error triggers the growing mode, while
going backwards from the WKB form (\ref{WKB}) of $x_-$ results in a
mixture of (\ref{Besselxplus}) and (\ref{Besselxminus}) as
$\hat\tau\to 0$. Shooting from both sides would be required, but we
have not done this.




\begin{thebibliography}{99}
\bibitem{footnote} Ref.~\cite{Sperhake} quotes a maximum fraction of
  energy radiated of $23\%$, while its Fig.~2 shows a maximum number
  of $1.7$ orbits; we have divided the former by the latter. For a
  boost of $2.9$, the maximal energy radiated went up to $35\%$, but the
  maximal number of orbits is not given.
\end{thebibliography}
\end{document}